\documentclass[apj]{emulateapj} 

\usepackage{graphicx}
\usepackage{amsmath} 
\usepackage{txfonts}
\usepackage{hyperref}

\newcommand{\de}{\partial}
\newcommand{\msun}{M$_{\sun}$}
\newcommand{\nut}{\nu_\mathrm{t}}
\newcommand{\nuout}{\nu_\mathrm{out}}
\newcommand{\nus}{\nu_\mathrm{s}}
\newcommand{\omegak}{\omega_{\textsc{k}}}

\newcommand{\rin}{r_{\mathrm{in}}}
\newcommand{\rrl}{R_{\mathrm{RL}}}
\newcommand{\rout}{r_\mathrm{out}}
\newcommand{\rs}{r_\mathrm{s}}
\newcommand{\rsun}{R$_{\sun}$}
\newcommand{\texp}{t_{\mathrm{exp}}}
\newcommand{\tmax}{t_{\mathrm{max}}}

\newcommand{\vsound}{\mathrm{v}_\mathrm{sound}}
\def\aap{A\&A}%
\def\apjl{ApJ}%
\def\apj{ApJ}%
\def\apjs{ApJS}%
\def\mnras{MNRAS}%
\def\araa{ARA\&A}%
\def\nat{Nature}%
\def\pasj{PASJ}%
\def\nar{New Astronomy Rev.}%
\def\actaa{Acta Astronomica.}%

\makeindex
\citeindextrue
\received{February 8, 2015}
\accepted{March 30, 2015}
\published{May 5, 2015}
\journalinfo{Astrophysical~Journal}
\submitted{}

\title{Evolution of finite viscous disks with time-independent viscosity}
\shortauthors{Lipunova}

\begin{document}

\title{Evolution of finite viscous disks with time-independent viscosity}

\author{G. V. Lipunova}
\affil{Lomonosov Moscow State University, Sternberg Astronomical Inst., 
Universitetski pr. 13, Moscow 119991, Russia; galja@sai.msu.ru}


\begin{abstract}
We find the Green's functions for the accretion disk with the fixed outer radius
and time-independent viscosity. With the Green's functions, a viscous evolution
of the disk with any initial conditions can be described. Two types of the inner
boundary conditions are considered: the zero stress tensor and the zero
accretion rate. The variable mass inflow at the outer radius can also be
included. The well-known exponential decline of the accretion rate is a part of
the solution with the inner zero stress tensor. The solution with the zero
central accretion rate is applicable to the disks around stars with the
magnetosphere's boundary exceeding the corotation radius. Using the solution,
the viscous evolution of disks in some binary systems can be studied. We apply
the solution with zero inner stress tensor to outbursts of short-period X-ray
transients during the time around the peak. It is found that for the Kramers'
regime of opacity and the initial surface density proportional to the radius,
the rise time to the peak is $t_\mathrm{rise} \approx 0.15\, \rout^2/\nuout$ and
the $e$-folding time of the decay is $t_\mathrm{exp} \approx 0.45
\,\rout^2/\nuout$. Comparison to non-stationary $\alpha$-disks shows that both
models with the same value of viscosity at the outer radius produce similar
behaviour on the viscous time-scale. For six bursts in X-ray novae, which
exhibit fast-rise-exponential-decay  and are fitted by the model, we find
a way to restrict the turbulent \hbox{parameter~$\alpha$}.
\end{abstract}

\keywords{accretion, accretion disks -- binaries: general -- methods: analytical}

\section{Introduction}
\label{s.intro}

The accretion processes are recognized to power many astrophysical
sources,  which are observable due to the release of energy when
matter is spiralling down  into the gravity well.  The study
of viscous accretion disks as underlying mechaisms for the extraction
of potential gravitational energy began with the works
of~\citet{lyndenbell1969,sha-sun1973}, \citet[pp.343-450]{novikov-thorne1973}.

Time-dependent problems for the accretion disks arise when describing a
wide variety of outburst phenomena, observed in the accreting sources,
i.e. binary systems with mass transfer and galactic nuclei, and systems
with planet formation. It is usually possible to consider separately 
the vertical and radial structure of an accretion disk due to
significantly different characteristic time-scales. In such a case the
time-dependent radial structure of a disk can be described by a
second-order partial differential equation, which is a consequence of
the angular momentum and mass conservation. The energy conservation in a
geometrically thin optically thick disk is provided by the local balance
of the viscous heating and radiative cooling.

The basic equation of the viscous evolution,
Eq.~\eqref{eq:time-dep-acc1} or~\eqref{eq:basic_tda_Kepler} below,
describes the viscous angular momentum flux along the disk radius. This
is a diffusion type equation with a variable `diffusion' coefficient.
Choice of an analytic approach to solve the equation depends on the
nature of the viscosity involved. The  preferential way of
describing viscosity in as\-tro\-phy\-sical disks, although not
exclusive, is to use \hbox{$\alpha$-viscosity}~\citep{shakura1972,
sha-sun1973}, i.e., the proportionality of the viscous stress tensor to
the total pressure in the disk. One can generally assume that the
kinematic viscosity $\nu$ is a product of two power functions: one
function of radius and another of some local physical parameter, the
latter usually represented by the surface density.

The boundary conditions imposed on the accretion disk are important.
While the outer boundary can be a freely expanding surface due to the
spreading of the matter corresponding to the  outward viscous
angular momentum flux, in some situations certain conditions should be
posed at some distance from the accreting object. Disks in the binary
systems are the main focus for such models.

In a binary system, the tidal torques, acting inside the Roche lobe of
the accreting object, truncate the disk and provide a sink for the
angular momentum. \citet{pap-pri1977} show that the effective tidal
radius is close to the Roche lobe: $\sim 0.9$ of the mean radius of the
Roche lobe, well coinciding with the largest non-intersecting periodic
orbits in the  restricted three-body problem \citep{paczynski1977}.
Numerical models confirm that most of the tidal torque is applied in a
narrow region at the edge of the disk, where perturbations become
nonlinear and strong spiral shocks appear~\citep[see][and references
therein]{pringle1991,ich-osa1994,hameury-lasota2005}. In the present
study, we leave aside the outer disk radius variations and other
manifestations of the tidal interactions in the binary except for the
disk truncation.  We confine ourselves to searching for a solution
to a problem of the viscous evolution of an accretion disk with the
constant outer radius. 

We investigate a geometrically thin, axisymmetric Keplerian finite
accretion disk with a viscosity in the form $\nu \propto r^b$, which is
time-independent. We obtain Green's functions (\ref{eq:green_zeroBi}),
(\ref{eq:Mdot_green_zeroBi}), (\ref{eq:green_zeroAi}), and
(\ref{eq:Mdot_green_zeroAi}), which are the kernels in the integrals
(\ref{eq:f_from_green}), (\ref{eq:mdot_from_green}), and
(\ref{eq:sol_nonzeroMdot_full}) to compute the viscous angular momentum
flux\footnote{The couple $g$ of~\citet{lyn-pri1974}.} $F$ and the
accretion rate $\dot M$. The method of Green's functions allow one to
compute $\Sigma(r,t)$ and $\dot M(r,t)$ for arbitrary initial surface
density  distributions. Different boundary conditions are considered: no
stress or no accretion at the inner boundary, which is located at the
zero radial coorinate, zero or time-dependent mass inflow at the outer
boundary. 

It is beleived that the mechanism of bursts in X-ray novae involves the
thermal instability connected  with opacity variations in the
disks, as in dwarf novae. When a burst is ignited, and a sufficient
portion of the disk is in the hot state, the viscous diffusion starts to
govern the evolution, driving the characteristic rise and drop of the
accretion rate onto the central object. It is a question,  which
probably allow no universal answer,  why X-ray novae light curves have 
different profiles and pecularities. If the disk is not very large, as
it happens in short-period X-ray transients, its whole body is likely to
be engaged in the viscous evolution. It has been shown that an episodic
mass input at the outer radius causes the fast-rise-exponential-decay
(FRED) profile due to the
viscous evolution~\citep{wood_etal2001}. Using the Green's function, one
can explore the situation with an arbitrary initial distribution of the
matter. Actually,  as we show in the present study after obtaining
the Green function, for realistic initial density distributions and for
the case with the zero inner viscous stress, the evolution  is
ge\-ne\-ral\-ly a FRED. It is interesting to compare the results of the
model with what happens in the $\alpha$-disks, the prevalent model for
the disks' turbulence in the stellar binaries. 

The second case, when the viscous stress is non-zero at the center,
applies to the disks with the central source of the angular momentum,
e.g., rotating stars with magnetosphers. If the magnetosphere's radius
is large enough, the matter has a super-Keplerian angular momentum at
its boundary. Different regimes for such disks have been proposed:
propellers and `dead' disks. Using the Green founction, the disks of
constant mass that evolve to dead disks can be described. 

In binary systems the mass transfer rate is a key factor. At some stages
it can be considered negligible comparing to the accretion rate in the
disk. In other systems, mass transfer variations may cause the activity 
in the disk. The case of time-dependent mass transfer, which includes
the constant transfer rate as a particular case, can be also studied
analytically  if $\nu\propto r^b$. Using the Green's functions, it
is possible to express $\Sigma(r,t)$ and $\dot M(r,t)$ for any outer
boundary conditions (for any $\dot M_\mathrm{out} (t)$) as well as for
any initial conditions.

In Section 2, we present the viscous evolution equation and review its
known solutions for accretion disks. In Section~\ref{sec:ev_fin_disk},
the Green's functions are found for different inner and outer boundary
conditions. An analytic way to calculate the disk evolution with the
variable mass transfer at the outer radius is presented; the features of
a radiating dead disk are considered. In Section~\ref{s.decay}, a stage
of decaying accretion is dealt with and a comparison to the
$\alpha$-viscosity models is made. Application to the FRED light curves
of short-period X-ray transients is presented in 
Section~\ref{s.xnovae}. The last sections are dedicated to the
discussion and summary.

\section{Viscous accretion disk equation}
\label{s.equation}

The  equation of the viscous evolution in the accretion disk is
the diffusion-type equation \citep[e.g.,][]{kfm_book1998}:
\begin{equation}
\frac{\partial \Sigma} {\partial t} =  
\frac{1}{r}\, \frac{\partial}{\partial r} \left[ 
\frac{1}{\partial (\omega \, r^2)/\partial r} \,
\frac {\partial}{\partial r} (W_{r\varphi}\,r^2)
\right]\,  ,
\label{eq:time-dep-acc1}
\end{equation}
where, among the standard designations of the time, radius, angular
speed, and the surface density we have the component of the viscous
stress tensor $w_{r\varphi}$, integrated over the full thickness of the
disk
$$
W_{r\varphi}(r,t) =2 \int\limits_0^{z_\mathrm{o}}w_{r\varphi}
\,  \mathrm{d}z~,
$$
where $z_0$ is the half-thickness of the disk at  radius $r$, and the
stress tensor is related to the differential rotation of gas masses by
the kinematic viscosity $\nu$ as follows:
\begin{equation}
w_{r\varphi}
= - \rho\, \nu\, r\, \frac{\mathrm{d} \omega}{\mathrm{d} r}\, .
\label{eq:stress_disk}
\end{equation}
For the Keplerian disks $\mathrm{d}\omega/\mathrm{d}r=-(3/2) \omegak/r$,
and we rewrite
\begin{equation}
W_{r\varphi}(r,t) =  3\, \omegak\int\limits_0^{z_\mathrm{o}}
\nu\, \rho\, \mathrm{d}z~.
\label{eq:stress_kepl_int}
\end{equation}
With the formula for the surface density
\begin{equation}
\Sigma (r,t) = 2\, \int\limits_{0}^{z_0} \rho(r,z,t) \, \mathrm{d} z \, ,
\label{eq:surface_density}
\end{equation}
 assuming that $\nu$ is independent on $z$, we have
\begin{equation}
W_{r\varphi}(r,t)
= \frac 32\, \omegak\,\nu\, \Sigma\, .
\label{eq:stress_kepl_av}
\end{equation}

It is convenient to introduce a new independent variable
\hbox{$h(r)=v_\varphi (r)\, r = \omega\, r^2$} -- the specific angular
momentum. Then, for the Keplerian disk, equation~(\ref{eq:time-dep-acc1})
can be rewritten as:
\begin{equation}
\frac{\de \Sigma}{\de t}\, = \,\frac 34\, \frac{(G\,M)^2}{h^3}\,
\frac{\de^2 (\Sigma \,\nu\, h)}{\de h^2}\, ,\qquad h\equiv h_\mathrm{K}~.
\label{eq:basic_tda_Kepler}
\end{equation}

The acknowledged approach is to solve the disk evolution equation for
the variable proportional to the full viscous torque acting between the
adjacent rings in the disk, $\omega_{r\varphi}\cdot z\,2\pi r \cdot r$.
This way, the boundary conditions are easy to express. From the above
expressions, the following relation between $F=2\,\pi\,W_{r \varphi}r^2$
and the accretion rate can be obtained:
\begin{equation}
- 2\,\pi\, \Sigma\, v_r\, r~\equiv \dot M (r,t) =
\frac{\partial F}{\partial h}~,
\label{eq:acc-rate2}
\end{equation}
where the radial velocity $v_r$ is negative. 

One chooses a procedure to solve (\ref{eq:basic_tda_Kepler}) provided
the form of $\nu=\nu(r,\Sigma)$ is known. This form is determined by the
physical conditions in the disk. For example, the vertical structure can
be solved for the $\alpha$-disk, and the required relation between $\nu$
and $\Sigma$ can be found. Let us give a perspective of some studies
(mainly analytic) of the viscously evolving accretion disks. They differ
in respect of the viscosity prescription and boundary conditions. The
analytic solutions are the benchmarks for more sophisticated and
involved numerical models.

\subsection{Viscosity $\nu \propto r^b$}  

For the kinematic viscosity in the form $\nu \propto r^b$, the equation
of the viscous evolution~\eqref{eq:basic_tda_Kepler} becomes a linear
partial differential equation. In \citeyear{lust1952}
\citeauthor{lust1952} obtained particular solutions to the equation
proposed by his teacher von~Weizs\"acker~(\citeyear{weizsaecker1948})
and described the principles to determine the general solution for the
infinite and finite problems. 

For the accretion disks that can extend {\em infinitely},
\citet{lyn-pri1974} (hereafter LP74) find, for the two types of the
inner boundary conditions, the Green's functions that provide means
to describe the whole development of the disk. The inner radius of the
disk is zero in their exact solution. The long-term self-similar
evolution of the accretion rate goes as a power-law $\dot M \propto
t^{-(l+1)} $ with a parameter $l<1$ (see also Appendix A).

 \citet{pringle1991} considers the properties of an infinite  accretion
disk with a central source of the angular momentum using the 
Green-function technique. A related problem is solved by
\citet{tanaka2011} who finds the Green's function for an infinite disk
with the boundary conditions posed at the finite inner radius.

 \citet{kin-rit1998} consider a {\em finite} disk with the viscosity
constant over the radius and in time and derive the exponential decay of
the accretion rate in the disk.

 \citet{zdziarski_etal2009} study the mass flow rate through a disk
resulting from a varying mass-supply rate, derive the Green's function
for the accretion-rate and its Fourier transform. By means of numerical
simulations, they investigate a {\em finite} disk, residing between
non-zero $r_\mathrm{in}$ and $r_\mathrm{out}$, with the non-zero
accretion rate at the outer radius. \citet{tanaka_et2012} find general
Green's function for an explicit dependence of the inner boundary
$r_\mathrm{in}$ on time and a non-zero mass across $r_\mathrm{in}$.

 In the present work, we study analytically a finite accretion disk with
the zero $r_\mathrm{in}$ and find an analytic solution describing its
whole evolution, from the rise to decay. For the case of the zero stress
tensor in the disk center and the mass deposition at the outer
disk edge, the Green's function of the problem is found in
\citet{wood_etal2001}.

\subsection{Viscosity $\nu \propto \Sigma^a r^b$}

Similarity solutions of the first kind of the non-linear differential
equation~\citep{barenblatt1982e} have been found at the
"decaying-accretion stage," when the total angular momentum of an {\em
infinite} disk is constant~\citep{pringle1974PhD,pringle1981}. Accretion
rate is found to decrease as $\propto t^{-5/4}$ for the Thomson opacity,
and as $\propto t^{-19/16}$ for the Kramers' opacity~\citep[see
also][]{filipov1984, cannizzo_etal1990}. \citet{lyub-shak1987} obtain
self-similar solutions that describe separately three stages of the disk
evolution. The first two are the self-similar solutions of the second
kind: the initial movement of the inner edge of a ring of matter toward
the center and the rise of the accretion rate in the center ($\dot M
\propto t^{2.47}$ for the Kramers' opacity and $ \propto t^{1.67}$ for
the Thomson opacity). The third stage is the decay of the accretion and
corresponds to the solution of \citet{pringle1974PhD}. The first two
stages of \citet{lyub-shak1987} may be applicable to the {\em finite}
disks when the conditions at the distant outer boundary do not affect
the behaviour of the disk near the center.

\citet{lin-pringle1987} derive an analytic self-similar solution for a
disk subject to the gravitational instability, which transfers mass and
angular momentum in the disk, for which $\dot M \propto t^{-6/5}$, and a
numerical solution for the whole evolution.  \citet{lin-bodenheimer1982}
found a self-similar solution for $\nu \propto \Sigma^2$, if the
viscosity is due to the action of convectively driven turbulent viscous
stresses ($\dot M \propto t^{-15/14}$).

Accretion disks with a central source of angular momentum and no mass
transfer through the inner boundary were considered by
\citet{pringle1991} who derived a self-similar solution (see also
\citet{ivanov_etal1999}). \citet{rafikov2013} in the extensive study of
the circumbinary disks around supermassive black holes obtained
self-similar solutions for the disks with a possible mass transfer
across the orbit of the secondary black hole.

For a {\em finite} disk, the radial and temporal solution is obtained by
\citet{lip-sha2000} yielding $\dot M \propto t^{-10/3}$ for the Kramers'
opacity and $\dot M \propto t^{-5/2}$ for the Thomson opacity.

In the context of the self-similarity solutions, we would like to note
that for an advection-dominated infinitely expanding accretion flow,
\citet{ogilvie1999} obtains by similarity methods a time-dependent
solution with the conserved total angular momentum.

\section{Evolution of the finite disk with steady viscosity}
\label{sec:ev_fin_disk}

If the kinematic viscosity $\nu$ is not a function of $\Sigma$ and is a
function of the radius alone, we arrive at the linear differential
equation~\eqref{eq:basic_tda_Kepler}. We write the kinematic
viscosity as
\begin{equation}
\nu = \nu_0\, r^b~.
\label{eq:nut}
\end{equation}
Then Eq.~(\ref{eq:basic_tda_Kepler}) for the dependent variable $F$
is as follows
\begin{equation}
\frac{\partial F}{\partial t}= \frac 34\, { \nu_0\,h^{2b-2}\,
(G\,M)^{2-b}}\,\frac{\partial^2F}{\partial h^2}~,
\label{eq:lin-diff}
\end{equation}
where
\begin{equation}
F = 2\, \pi\,W_\mathrm{r\varphi}\,r^2 = 3\,\pi\, h\,  \Sigma\,\nu_0\, r^b\, .
\label{eq:F_sigmo}
\end{equation}
We rewrite \eqref{eq:lin-diff} in the form similar to that in LP74:
\begin{equation}
\frac{\partial^2F}{\partial h^2} = \frac 14\,
\left(\frac{\kappa}{l}\right)^2\, h^{1/l-2}\, \frac{\partial F}{\partial
t}\, ,
\label{eq:diff_lyn-pringle}
\end{equation}
where the constants are defined as follows:
\begin{equation}
\frac{1}{2\,l} = 2-b\, , \qquad \kappa^2 = \frac{16 \, l^2}
{3\nu_0\,(G\,M)^{1/2l}}\, .
\label{eq:consts_lb}
\end{equation}
There is a degeneracy of the index $l$ at $b=2$. But it is only apparent
as $l$ enters equation (\ref{eq:diff_lyn-pringle}) in the denominators.
Value $b=2$ corresponds to the change of the class of functions, which
satisfy the differential equation, and is left out of the following
consideration. In the case of $b=2$ the viscous time is constant over
the disk radii $t_\mathrm{vis} \propto r^2/\nu_o r^b$. We limit
ourselves to the finite positive values of $l$ and the Keplerian disks.

In the astrophysical hot accretion disks, the turbulent viscosity is at
work. According to the Prandtl concept~\citep{prandtl1925}, the
kinematic turbulent viscosity is an averaged product of the two random
values: $\nut=\overline{l_\mathrm{t}\,\upsilon_\mathrm{t}}$, where
$l_\mathrm{t}$ and $\upsilon_\mathrm{t}$ are the path and speed of the
turbulent motion. For the component of the viscous stress tensor
we write $w_{r\varphi} = - \rho\, \nut\, r \,{\mathrm{d}
\omega}/{\mathrm{d} r}$.
Following the Prandtl consideration,
\hbox{$\upsilon_\mathrm{t} = -l_\mathrm{t}\, r\, {\mathrm{d}
\omega}/{\mathrm{d} r}$}, which leads to \hbox{$w_{r\varphi} = \rho\,
\overline{\upsilon_\mathrm{t}^2}$}. 
\citet{shakura1972} proposed the $\alpha$-disks:
$w_{r\varphi} = \alpha\, \rho\, \upsilon_\mathrm{s}^2$. 
From the hydrostatic balance one derives the
sound speed $\upsilon_\mathrm{s} \approx \omega\, r\ (z_0/r)$ and
obtains for a Keplerian disk $\nut \approx 2/3 \, \alpha\,h\,
(z_0/r)^2$. If the relative half-thickness of the $\alpha$-disk is
invariable then $b=1/2$ and $l=1/3$. This value of $b$ takes place for
the `$\beta$-viscosity disks' with $\nu =
\beta\, h$, suggested by \citet{duschl_et2000} to act in the
protoplanetary and galactic disks. The
standard model of the $\alpha$-disk implies  that $h/r$ varies
with $r$ and $\Sigma$.

Let us find the Green's function of the linear 
equation~(\ref{eq:diff_lyn-pringle}) with the specific boundary
conditions at the boundaries. 
A Green's function or a diffusion kernel is a reaction of the system to
the delta impulse, in other words, the Green's function is the solution
to the differential equation with the initial condition $F(x,0) =
\delta(x-x_1)$, where $\delta(x-x_1)$ is the Dirac delta function. We
search a solution by separating the variables, \hbox{$F(h,t)=
f(h_\mathrm{out} \, \xi) \, \exp (-s\,t)$}, where $s$ is a constant and
$h_\mathrm{out}$ is the outer disk radius. Substituting this into
(\ref{eq:diff_lyn-pringle}), we get
\begin{equation}
\frac{\mathrm{d}^2f(\xi)}{\mathrm{d} \xi^2} + \frac {1}{4}\,
\left(\frac{k}{l}\right)^2\, \xi^{1/l-2}\, f(\xi) =0\, , 
\label{eq:sturm-liuv}
\end{equation}
 where $k^2 =
\kappa^2\, h_\mathrm{out}^{1/l} \, s $. This is a Lommel's transformation of 
the Bessel equation, which can be derived
by introducing a new independent variable $x = \xi^{1/2l}$.
A particular solution is represented as
$$
F_k(x,t) = e^{-st}\, (kx)^l \, [A\,J_l(kx) + B\, J_{-l}(kx)]\, 
$$
\citep{lust1952}, where $J_l$ and $J_{-l}$ are the Bessel functions of a
non-integer order. If $l$ is an integer number, the Bessel functions of
the second kind should be used instead of $J_{-l}$. For $b=2$, the
solution is no more a Bessel function~\citep[see Appendix I
of][]{kfm_book1998}. The value $b=2$ marks a situation when the
characteristic viscosity time $ r^2/\nu$ is constant over the radius.
We do not consider this case in the present study.

A general solution for an infinite
problem is a superposition of par\-ti\-cu\-lar solutions and is
expressed by LP74 as an integral over all positive values of $k$:
$
F(x,t) = \int_0^\infty F_k(x,t)\, \mathrm{d}k\, .
$
For a problem inside a finite interval, a general solution is a
superposition of particular solutions and is expressed as an infinite
sum~\citep{lust1952}:
\begin{equation}
F(x,t) = \sum_{i=1}^\infty e^{-t\,k_i^2\,\kappa^{-2}\, h_\mathrm{out}^{-1/l}} \,
(k_i\,x)^l \, [A_i\,J_l(k_i\,x) + B_i\, J_{-l}(k_i\,x)]\, ,
\label{eq:gen_sol_finite}
\end{equation} 
where constants $k_i$, $A_i$. and $B_i$ are to be defined from two 
boundary conditions and one initial condition. The dimensionless
coordinate $x$ lies in the interval $[0,1]$, while the physical
coordinate $h$ (specific angular momentum) lies inside
$[0,h_\mathrm{out}]$. The outer boundary condition, which sets the
accretion rate at the outer radius of the disk, can be written as
follows
\begin{equation}
\frac{\partial F}{\partial h} = \dot M_\mathrm{out}(t) 
\mbox{~~at~~} h=h_\mathrm{out}~. 
\label{eq:zero_dotM_out}
\end{equation}
In the simplest case, the homogeneous outer-boundary condition
corresponds to the zero mass-inflow rate at the outer boundary.
As this happens, the viscous angular momentum flux $F$ takes some non-zero
value, which implies  withdrawal of the angular momentum at the
outer boundary. This corresponds to the situation of the disk in the
binary system, when the angular momentum is removed from the outer edge
of the disk  by the tidal action of the secondary.

At \hbox{$x=0$}, one trivial inner boundary condition \hbox{$F(x,t)=0$}
approximates the structure for the so-called `standard' disk, for which 
the absence of the viscous stresses at the disk inner edge is assumed.
Another simple case, $\partial F/\partial h=0$ at $x=0$, describes the
situation of no accretion through the inner disk edge. This case is
applicable for the stars with strong magnetosphere, when the Alfv\'en
radius is greater than the corotation radius, impeding the accretion of
the matter. We consider these two cases separately.

\subsection{The case of  zero viscous stress at the inner
boundary and zero accretion rate at $r_\mathrm{out}$}
\label{sec:no_visc}

First, we consider the case of the zero viscous stress at the inner
boundary. Approximately, this corresponds to the solution $F(x,t)=0$ at
$x=0$ and only the coefficients at $J_l$-terms in
(\ref{eq:gen_sol_finite})
remain non-zero.  Thus, sum (\ref{eq:gen_sol_finite})
at $t=0$ becomes
\begin{equation}
F(x,0) = \sum_{i=1}^\infty 
(k_i\,x)^l \, A_i\,J_l(k_i\,x) \, .
\label{eq:gen_sol_finite_zeroBi_t0}
\end{equation}

We assume the zero accretion rate at $r_\mathrm{out}$. Such boundary
condition approximates the situation of the low mass-transfer from the
se\-con\-da\-ry in the binary system, for example, when the accretion
rate in the disk is much higher than the mass transfer rate during an
outburst. Dropping the time part of the solution, we arrive at 
condition (\ref{eq:zero_dotM_out}) required for every term in the sum
(\ref{eq:gen_sol_finite_zeroBi_t0}), which can be rewritten as follows:
\begin{equation}
l \, J_l (k_i) + k_i\, J_l' (k_i) = 0\,.
\label{eq:out_boundary}
\end{equation}

Thus, the general solution for the case of zero viscous stress tensor at
the center and zero accretion rate at the outer boundary is the sum
(\ref{eq:gen_sol_finite}), where $B_i=0$ and $k_i$ are the positive
roots of the transcendental equation (\ref{eq:out_boundary}). To define
$A_i$ from the initial condition, the finite Hankel transforms are used.

The series $ \sum\limits_{i=1}^\infty \, k_i^l\,A_i\,J_l(k_i\,x) $ with
condition (\ref{eq:out_boundary}) are known as the Dini's series
\citep[see][\S 18.11]{watson1944treatise}. Function $f(x)=F(x,0)\,
x^{-l}$ can be expressed as the Dini's expansion if the function
satisfies the Dirichlet conditions in the interval and the coefficients
are defined as $k_i^l\, A_i = 2\, \bar f_J(k_i) \, J_l^{-2}(k_i) $
\citep{watson1944treatise,sneddon1951fourier} with
the finite Hankel transform 
$$
\bar f_J(k_i) =  \int\limits_0^1 x\, f(x) \, J_l(k_i\, x)\, \mathrm{d} x~,
$$
where $f(x) = F(x,0)\, x^{-l}$. We look for a solution
for the $\delta$-function as the initial condition:
$F(x,0)=\delta(x-x_1)$. Using the properties of the
Dirac function, we obtain
\begin{equation}
k_i^l\,A_i = 2 \, x_1^{1-l}
\frac{J_l(k_i\,x_1)}{J_l^2(k_i)}\,.
\label{eq:Ai}
\end{equation}

\begin{figure}
\centering
\includegraphics[angle=0,width=0.45\textwidth]{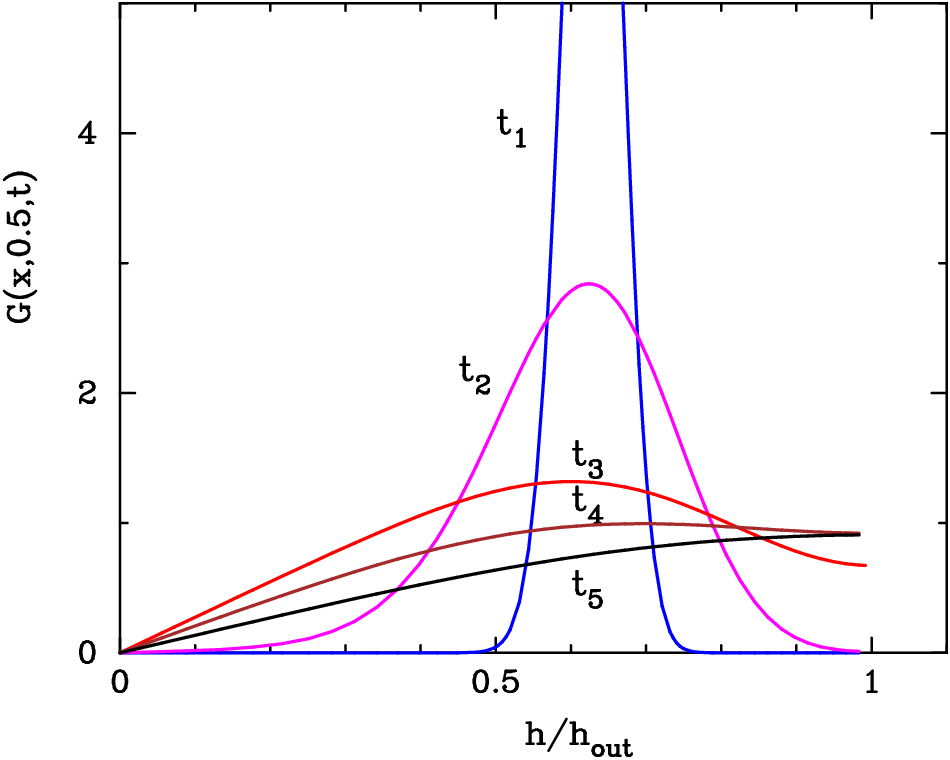}
\caption{Green's function for the finite accretion disk with the  zero 
viscous stress in the center  at 
five moments of time $t_1=0.001$, $t_2=0.01$, $t_3=t_{max}^\infty=3/64$,  
$t_4=0.1$,
$t_5=0.3$. Initial spike was
at position $x_s=(h/h_\mathrm{out})^{1/2l}=0.5$. Constants  $\kappa=1$,
$l=1/3$.
\label{fig.green_finite}
}
\end{figure}

Thus, for the specific boundary conditions, we derive the Green's
function, which is the solution to (\ref{eq:diff_lyn-pringle}) with
$\delta$-function as the initial distribution
\begin{equation}
G(x,x_1,t) = 2 \, x^l\, x_1^{1-l}\,
 \sum_i e^{-t\,k_i^2\,\kappa^{-2}\, h_\mathrm{out}^{-1/l}} \, 
\frac{J_l(k_i\,x_1)\,J_l(k_i\,x)}{J_l^2(k_i)}\,,
\label{eq:green_zeroBi}
\end{equation}
where $k_i$ are the positive roots of Eq.~(\ref{eq:out_boundary}) and $x
= (h/h_\mathrm{out})^{1/2l}$. The function is plotted in
Fig.~\ref{fig.green_finite} for consecutive times. The 
curve for time $t_3$  corresponds
to the maximum accretion rate in the center. It is plotted at 
$t_3=t_\mathrm{max}^\infty$ 
given by \eqref{eq.t_max_inf} below.

\begin{figure*}
\centering
\includegraphics[angle=0,width=0.45\textwidth]{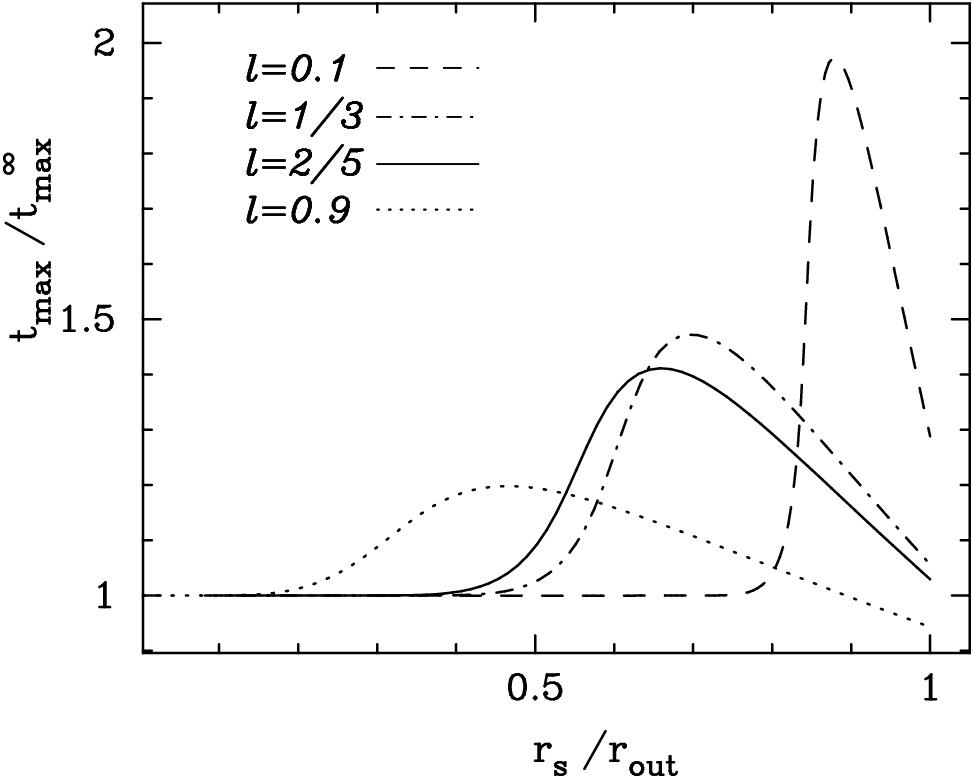}
\hskip 1cm
\includegraphics[width=0.45\textwidth]{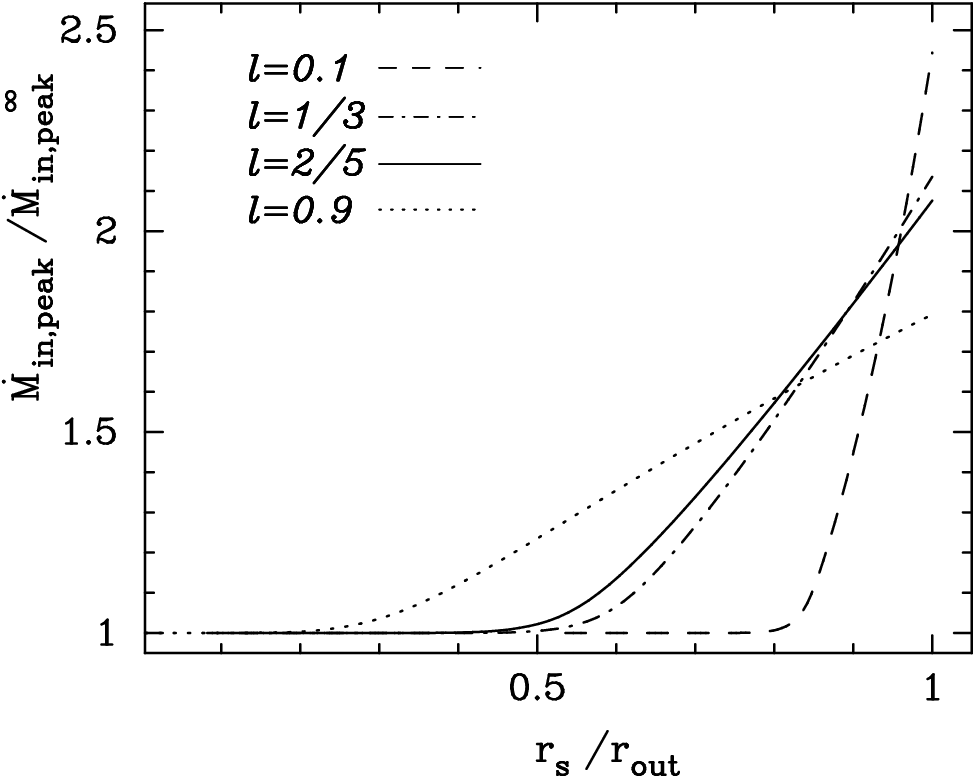}
\caption{
Left: ratio of the time when the accretion rate peaks in the finite
disk   to that in the infinite disk versus the position
of the initial thin ring. Right: ratio of the peak accretion rate in the finite
disk to that in the infinite disk as a dependence on the initial ring
position.
\label{fig.Cmax1}
}
\end{figure*}

Given the dimensional initial distribution $F(x,0)$ in the finite
accretion disk, the distribution of $F$ at any $t>0$ is
\begin{equation}
F(x,t) = \int\limits_0^1 F(x_1,0)\, G(x,x_1,t)\, \mathrm{d} x_1\,.
\label{eq:f_from_green}
\end{equation}
The accretion rate can be found from
\begin{equation}
\dot M (x,t) =\int\limits_0^1 F(x_1,0)\, G_{\dot M}(x,x_1,t) \, \mathrm{d}
x_1\Big/ {h_\mathrm{out}}  \,, 
\label{eq:mdot_from_green}
\end{equation}
where the Green's function for the accretion rate
\begin{equation}
\begin{split}
G_{\dot M}(x,x_1,t)& \equiv \frac{\partial 
G(x,x_1,t)}{\partial x^{2l}} 
= \\
&=\frac{(x\,x_1)^{1-l}}{l}  \sum_i e^{-t\,k_i^2\,\kappa^{-2}\, h_\mathrm{out}^{-1/l}} \,
k_i\,
\frac{J_l(k_i\,x_1)\,J_{l-1}(k_i\,x)}{J_l^2(k_i)}\,.
\end{split}
\label{eq:Mdot_green_zeroBi}
\end{equation}
Functions $G$ and $G_{\dot M}$  are found by~\citet{wood_etal2001} for
the case $x_1=1$ along with an analytical form obtained from the
asymptotic forms  for small and large $t$.

One can express the initial $F$, proportional to the disk mass, from the
initial distribution of the surface density, using
(\ref{eq:stress_kepl_av}) and (\ref{eq:consts_lb}):
\begin{equation}
F(x,0) = \frac{16\,\pi\,l^2}{\kappa^2\,h^{1/l}}\, r^2 \,
\Sigma(r)\, h
\label{eq:f_sigma_l}
\end{equation}
where $r=h^2/GM$ and $h= h_\mathrm{out} \,x^{2l}$.

\subsubsection{Spike-like and power-law initial distribution of the viscous stresses}

Let the initial mass be located at some radius $r_s$ where the Keplerian
specific angular momentum is $h_s = x_s^{2l}\,h_\mathrm{out}$, and the
mass of the infinitely thin ring is $M_\mathrm{disk}$, $h_\mathrm{out}$
is the maximum possible specific angular momentum of the matter.

The surface density of the $\delta$-ring is  
$$
\Sigma(r)  =\frac{M_\mathrm{disk}}{2\,\pi\,r_s}\,
\delta(r-r_s) = \frac {
M_\mathrm{disk}}{8\,\pi\,l\,r_s^2} \,
 x_s \,
\delta(x-x_s)\, .
$$
We take into account that $\delta(r-r_s)=\delta(x-x_s) \mathrm{d} x/
\mathrm{d}r$ and $r =x^{4l}\,r_\mathrm{out}$. Applying
(\ref{eq:f_sigma_l}), we obtain
\begin{equation}
F(x,0) =\frac{2\, l}{\kappa^2}\, M_\mathrm{disk}\,
h_\mathrm{out}^{1-1/l}\, x_s^{2l-1} \, \delta(x-x_s)\, 
\label{eq:Fring}
\end{equation}
and the accretion rate evolution is as follows
$$
\dot M (x,t) = \frac{M_\mathrm{disk}}{t_\mathrm{vis}}\,
x_s^{2l-1} \, G_{\dot M}(x,x_s,t)~,
$$
where we designate
$$
t_\mathrm{vis} = \frac{\kappa^{2}\, h_\mathrm{out}^{1/l}}{ 2\,l}
= \frac{8 \,l}{3}\, \frac{\rout^2}{\nu(\rout)} \, .
$$

The time $t_\mathrm{max}$ of the maximum and its accretion rate in
the disk center $\dot M_{\mathrm{in,peak}}$ are of  order of 
those for the infinite disks.  Using the solution of
LP74, we find the time of the accretion rate peak for the infinite disk
\begin{equation}
t_\mathrm{max}^\infty  =  \frac{\kappa^2\, h_s^{1/l}}{4\,(1+l)} = 
\frac{4\,l^2}{3\,(1+l)} \, \frac{\rs^2}{\nu(\rs)}
\label{eq.t_max_inf}
\end{equation}
and value  $\dot M^\infty_{\mathrm{in,peak}}$~\eqref{eq:ap.dotmmax_snfinite}.
The numerical factor in \eqref{eq.t_max_inf} is presented in
Table~\ref{tab:zeroes} for various, physically motivated  values of $l$.
\begin{figure*}
\centering
\includegraphics[width=0.99\textwidth]{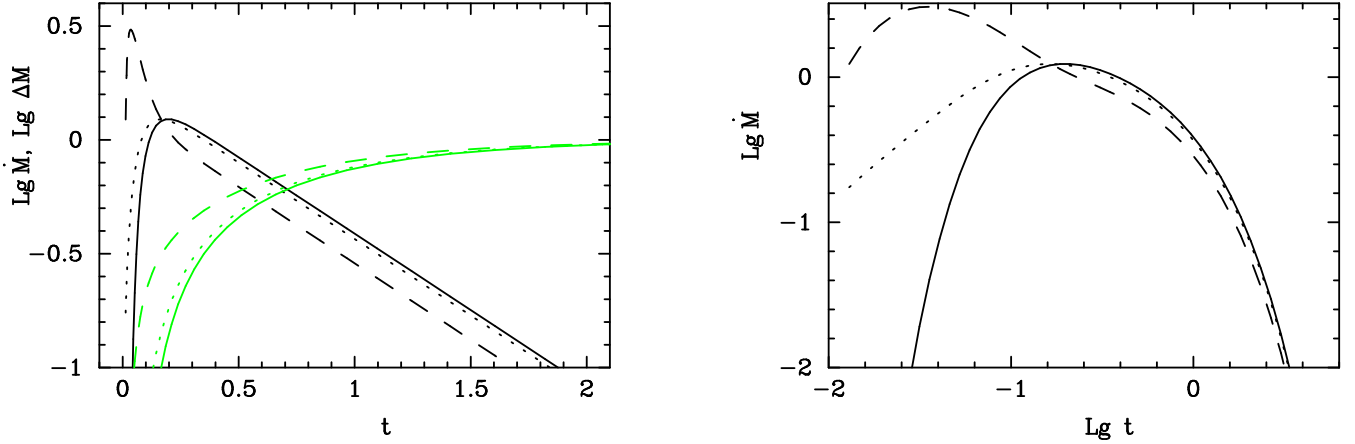}
\caption{The inner accretion
rate variation (the black curves with a peak) for three initial
distributions of $F(h)$ shown in Fig.~\ref{fig.g_init}. The initial mass
of the disk is the same in all cases. On the left panel, the cumulative
accreted mass is shown by the monotonic (green in the electronic
version) curves. The dashed curve has a power-law interval, which is 
clearly seen on the right panel. Constants $M_\mathrm{disk}=1$,
$h_\mathrm{out}=1$, $\kappa=1$, $l=1/3$, $t_\mathrm{vis}=3/2$.
\label{fig.mdotin}
}
\end{figure*}
In Fig.~\ref{fig.Cmax1} the ratios of maximum times and accretion rates
for the finite and infinite disk are shown for several cases of $l$. The
ratios approach $1$ for $x_s\ll 1$ -- the disk `does not feel' the
boundary at the time when the accretion rate peaks. 
\begin{figure}[b!]
\centering
\includegraphics[width=0.45\textwidth]{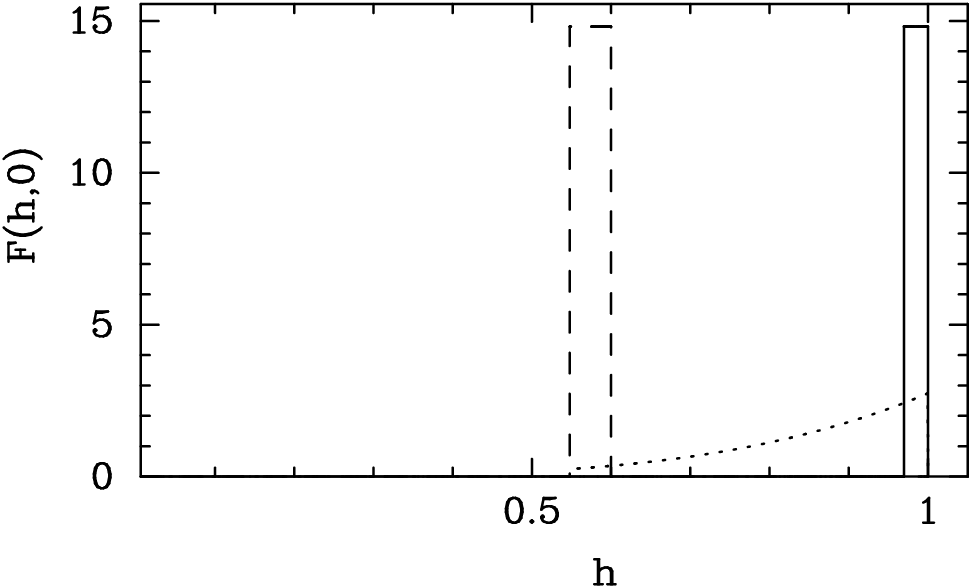} 
\caption{Three initial
distributions of $F(h)$ in the disk, whose inner accretion rate
evolution is shown in Fig.~\ref{fig.mdotin}. Top-hats are located at 
$0.94-1$ (solid line) and $0.3-0.36$ of $r_\mathrm{out}$ (dashed line)  and
the wide distribution, corresponding to $\Sigma(r)
\propto r$,  goes in $(0.1-1)\,r_\mathrm{out}$ (dotted line).
The disk mass is the same in all cases and equals 1. The outer specific
angular momentum $h_\mathrm{out}=1$. The top-hat distributions
are constructed using $\alpha_\Sigma=-1$. Parameter $l=1/3$.
\label{fig.g_init}
}
\end{figure}

If the  initial surface density is distributed as a power-law with radius
$\Sigma \propto r^{\alpha_\Sigma}$, we use the following expression in 
Eq.~(\ref{eq:f_sigma_l}):
$$
\Sigma(r) = \frac{M_\mathrm{disk}}{2\,\pi\,r^2} \, 
\frac{(\alpha_\Sigma+2) \, (r/r_2)^{\alpha_\Sigma+2}}{1-(r_1/r_2)^{\alpha_\Sigma+2}}~,
$$
assuming the mass of the disk is enclosed between the radii $r_1$ and
$r_2$. 

In Fig.~\ref{fig.mdotin} the variation of the accretion rate at the disk
center is presented for three different initial distributions of
$F$~(see Fig.~\ref{fig.g_init}): two narrow top-hats
($\alpha_\Sigma=1/2l-5/2$) with different $r_1$ and $r_2$ and the one
corresponding to $\Sigma(r,0) \propto r$. All three initial
distributions are constructed for the same initial mass of the disk
\hbox{$M_\mathrm{disk}=1$}. The evolution of a thin ring located far
from the outer disk edge has a characteristic power-law interval, when
the disk behaves as if it was infinite. The interval lasts about
$t_\mathrm{max} [1- (h_\mathrm{s}/h_\mathrm{out})^{1/l}]$. If
$\Sigma(r,0) \propto r$, the mass of the disk at $t=0$ is
enclosed mainly in the outer parts, because it grows as the cube of the
radius. As expected, the accretion rate's evolution after the peak
depends weakly on the peculiarities of the initial distribution of mass
in the disk (compare the dotted and solid line in
Fig.~\ref{fig.mdotin}), as long as its bulk was similarly positioned.

\subsubsection{Disk luminosity}

The viscous heating in the
disk (its half) is found from $F$:
\begin{equation}
Q_\mathrm{vis}  = \frac{3}{8\,\pi} \, \frac{F\, \omegak}{r^2}\, ,
\label{eq.Qvis}
\end{equation}
 which can be equated to $\sigma T_\mathrm{eff}^4$ assuming the local
energy balance. 

In the center of the disk, the quasi-stationary distribution develops
soon and the accretion rate is constant over radii: $F(t) = \dot M(t)\,
h$. The zone of constant $\dot M$ expands over time.

The viscous heating \eqref{eq.Qvis} with the quasi-stationary
distribution of $F$ diverges at the zero coordinate. In the real
astrophysical disks, the inner radius of the disk is a finite value
$\rin \neq 0$. Note that integrating \eqref{eq.Qvis} over the disk
surface from $\rin$ to $\rout$ also yields the incorrect result for the
total power released in the disk. (As a remark, the integration of
\eqref{eq.Qvis} gives the correct result for the case of the zero
accretion rate at the center, see \S~\ref{s.deaddisk}).

In fact, setting $F(\rin)=0$ at the finite radius $\rin$ leads to a
time-independent correction of $F$ near $\rin$. This correction is found
for the infinite disks by \citet{tanaka2011} and represents the
classical factor $(1-\sqrt{\rin/r})$. Naturally, the correction holds
for a finite disk. It follows that the bolometric luminosity can be
calculated via the classical formula $L_\mathrm{bol} = G\, M\, \dot
M/(2\rin)$ at times $t>t_\mathrm{vis} (\rin/\rout)^{2-b}$ with the first
term in the correction of order of 
$(\rin/\rout)\,(t_\mathrm{vis}/t)^{1/(2-b)}$ for earlier times. For the
disks in the stellar binary systems, usually, ratio $\rin/\rout <<1$,
and we are safe to calculate $L_\mathrm{bol}$ by the classical formula
with $\dot M = \dot M (0,t)$ found from \eqref{eq:mdot_from_green}.

\begin{figure*}
\centering
\includegraphics[width=0.45\textwidth]{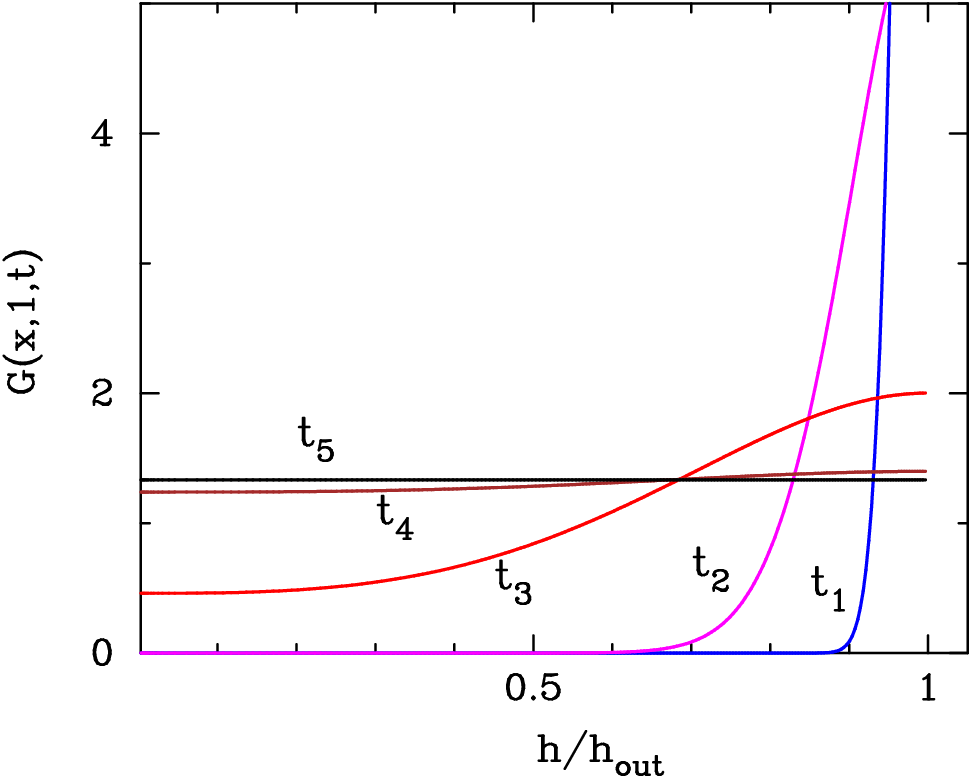}
\hskip 0.5cm
\includegraphics[width=0.45\textwidth]{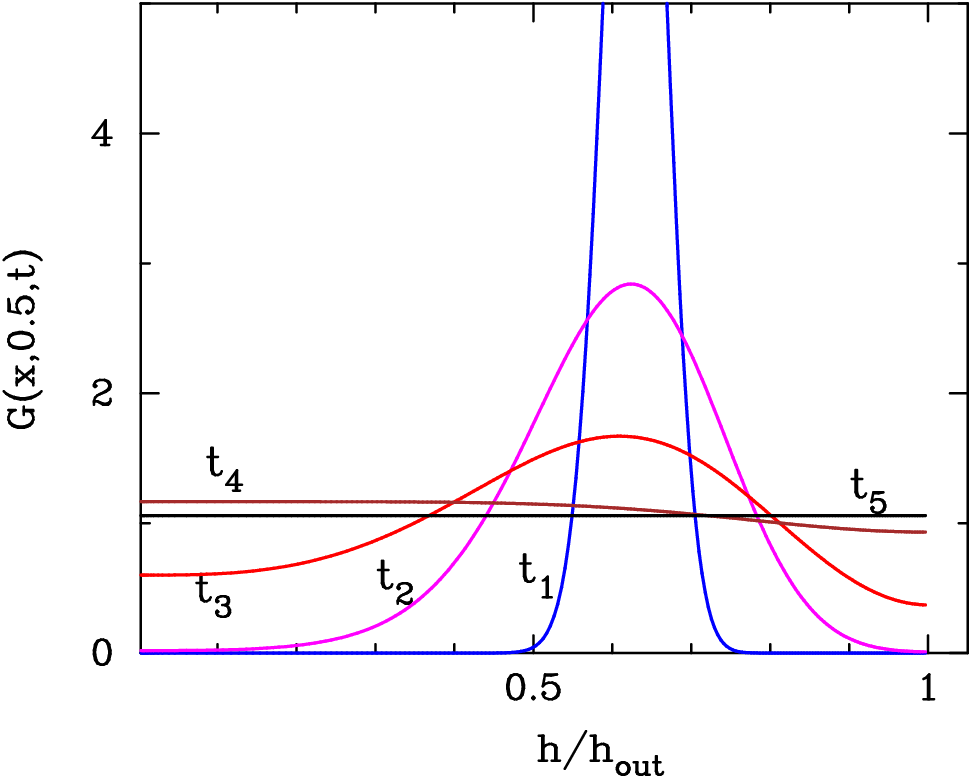}
\caption{Green's function for the finite accretion disk with the zero
accretion rate at the center at 
five moments of time $t_1=0.001$, $t_2=0.01$, $t_3=0.1$ $t_4=0.3$,
$t_5=0.8$ (left) and $t_1=0.001$, $t_2=0.01$, $t_3=0.03$ $t_4=0.1$,
$t_5=0.8$ (right). Initial spike is
at position $x_s=(h/h_\mathrm{out})^{1/2l}=1$ (left) and 0.5 (right). 
Constants  $\kappa=1$,
$l=1/3$, $t_\mathrm{vis} = 3/2$.
\label{fig.green_finiteB1}
}
\end{figure*}

\subsection{The case of the disk with no central accretion and zero accretion rate at $r_\mathrm{out}$} 
Condition $\partial F/\partial h=0$ at
\hbox{$h=0$} necessitates that all $A_i=0$ in \eqref{eq:gen_sol_finite},
which becomes for $t=0$:
\begin{equation}
x^{-l}\,F(x,0) =  B_0  + \sum_{i=1}^\infty 
k_i^l \, B_i\,J_{-l}(k_i\,x) \, ,
\label{eq:gen_sol_finite_zeroAi_t0}
\end{equation}
where $k_i$ are the positive roots. In the previous section, the zero
root contributed nothing to (\ref{eq:gen_sol_finite_zeroBi_t0}). Here 
the zero root provides a term like $(k_0\,x)^l J_{-l}(k_0\,x)$, which
must contribute something, because the Bessel function $J_{-l}$ behaves
as $\propto (k_0 \,x)^{-l}$ in the vicinity of zero. 

The method to find $B_i$ for $i\geq 1$ is analogous to that for $A_i$,
since the Dini's series are convergent for the order of the Bessel
function
$>-1$~\citep{benedek-panzone1972,guadalupe_etal1993}\footnote{The Hankel
integral transform was also proved for the Bessel orders
$>-1$~\citep{macrobert1932,Betancor-Stempak2001}.}. We consider only
non-integer $0<l<1$ and obtain
\begin{equation}
k_i^l \, B_i = 2 \, \, x_1^{1-l}
\frac{J_{-l}(k_i\,x_1)}{J_{-l}^2(k_i)}~.
\label{eq:Bi}
\end{equation}

It is known about the Dini's series of the negative order that an
additional first term is to be added to the sum, if the equation 
defining roots $k_i$ has particular coefficients\footnote{For the usual
form of writing the boundary condition for the Dini's series,
$H\,J_{\nu}(z) + z J'_{\nu}(z)$=0, the the additional term appears if
\hbox{$H+\nu=0$}~\citep{watson1944treatise}. }. Just this case occurs
for the properly rewritten outer boundary condition
(\ref{eq:zero_dotM_out}):
\begin{equation}
l \, J_{-l} (k_i) + k_i\, J_{-l}' (k_i) = 0~.
\label{eq:out_boundary_neg}
\end{equation}
The additional term in the Dini's series is expressed as
\begin{equation}
B_0 = 2\,(1-l)\, x^{-l}\int\limits_0^1 z^{1-l} \,F(z,0)\, z^{-l} \mathrm{d} z\,
,
\label{eq.B0}
\end{equation}
where $z$ is a free variable~\citep{watson1944treatise}. Thus, adopting
$F(z) = \delta (z-x_1)$, from \eqref{eq:gen_sol_finite} we obtain the
Green's function for the equation of the non-stationary accretion with
the zero inner accretion rate and the finite outer radius:
\begin{equation}
\begin{split}
G(x,x_1,t)& = 2 \,(1-l) \,x_1^{1-2l}\, + \\
&+\, 2 \, \left(\frac{x}{x_1}\right)^l\, x_1
\sum_{i=1}^{\infty} e^{-t\,k_i^2\,\kappa^{-2}\, h_\mathrm{out}^{-1/l}} \, 
\frac{\, J_{-l}(k_i\,x_1)J_{-l}(k_i\,x)}{J_{-l}^2(k_i)}~,
\end{split}
\label{eq:green_zeroAi}
\end{equation}
where $k_i$ are the positive roots of Eq.~(\ref{eq:out_boundary_neg})
and $x = (h/h_\mathrm{out})^{1/2l}$. Formulas (\ref{eq:f_from_green})
and (\ref{eq:mdot_from_green}) can be applied without changes.

This Green function is illustrated in Fig.~\ref{fig.green_finiteB1}. 
At the late stages, the accretion disks forget all information about the
initial distribution of the viscous stresses. For the disks without
accretion on to the center, the uniform distribution of $F$ develops,
with its magnitude being proportional to the mass stored in the disk.

We take the derivative of (\ref{eq:green_zeroAi}) and obtain the Green
function for the accretion rate:
\begin{equation}
G_{\dot M}(x,x_1,t) = \frac{x^{1-l}}{l} \, 
  \sum_i e^{-t\,k_i^2\,\kappa^{-2}\, h_\mathrm{out}^{-1/l}} \,
k_i\,
\frac{J_{-l}(k_i\,x_1)\,J_{1-l}(k_i\,x)}{J_{-l}^2(k_i)}~.
\label{eq:Mdot_green_zeroAi}
\end{equation}
It approaches zero when $x\rightarrow 0$ for any time $t$. This
corresponds to our boundary condition at the center: $\dot M =0$. 
This disk, without sink of mass and additional mass supply, 
emits radiation even when it reaches the `end' of its evolution, as the
viscous heating \eqref{eq.Qvis} does not stop when the flow of matter ceases.
If Fig.~\ref{fig.lumin_zeroAi} its bolometric luminosity is plotted
for the two initial positions of the ring (the details are given in the next
subsection).

\subsubsection{The dead disk}
\label{s.deaddisk}

The stable configuration of the confined `dead' disk (we adopt the
name from the work by \citealt{sunyaev-shakura77e}) has the following 
viscous  angular momentum flux at each $r$:
\begin{equation}
F_\mathrm{d} \equiv F|_{t\rightarrow\infty}  = \frac{4\,\, l \, (1-l)\, h_\mathrm{out}^{1-1/l}\,
M_\mathrm{disk}} { \kappa^2}
 = 2\,(1-l) \,\frac{h_\mathrm{out}\,
M_\mathrm{disk}}{t_\mathrm{vis}}\, ,
\label{eq:torque_dead}
\end{equation}
where \eqref{eq:Fring} is used and $t_\mathrm{vis} = 8\, l\,
\rout^{1/2l}\,(3\,\nu_0)^{-1}$ as before. 

The total luminosity of the dead disk is a constant value
$F_\mathrm{d}\,(\omega_\mathrm{in}-\omega_\mathrm{out})$. It can be
obtained by the integration of \eqref{eq.Qvis} over the disk surface
between $\rin$ and $\rout$. Unlikely to the case of the zero torque at
$r=\rin$, the late-time value of $F$ does not depend on the location at
which condition \hbox{$\dot M=0$} is set and solely depends on the
evolution at the large radii. This was shown for the infinite disks by
\citet{tanaka2011}. Value of $\rin$ affects the limunosity only by a
change in the size of the emitting surface. There is still a
time-dependent correction obtained by \citet{tanaka2011} having an
effect at early times $t<t_\mathrm{vis} (\rin/\rout)^{2-b}$.

The dead-disk state has to end sometime, as the disk cannot receive
infinitely the angular momentum from the central object. If the central
star has a magnetosphere, its radius $r_\mathrm{mag}$ can be estimated
from the equality of the viscous and magnetic torque at the
magnetosphere boundary~(LP74):
\begin{equation}
F(x\rightarrow 0) = \kappa_t
\frac{\mu_\star^2}{r_\mathrm{mag}^3}~,
\label{eq.rmag_cond}
\end{equation}
where $\mu_\star$ is the magnetic dipole of the star, $\kappa_t$ is a
dimensionless factor of order of unity~\citep{lipunov1992}. Within some
factor, this condition is equivalent to the equality of the gas and
magnetic pressure at the inner disk boundary~\citep{sunyaev-shakura77e}.
The accretion is inhibited if the inner disk radius, $r_\mathrm{mag}$,
is greater than the corotation radius, because the drag exerted by the
magnetic field is super-Keplerian:
$$
r_\mathrm{cor} = \left(\frac{G\, M_\star}{\omega_\star^2}
\right)^{1/3}\,  = 1.7\times 10^8\, P_{\star}^{2/3} \, M_{1.4}^{1/3}~\mbox{cm},
$$
where $M_\star$, $\omega_\star$ and $P_\star$ are the mass, angular
speed and period of the revolution of the central star. In the last
expression the mass is normalized by $1.4$~\msun. Thus we obtain for
a final steady configuration
$$
\frac{r_\mathrm{mag,d}}{r_\mathrm{cor}} \approx 1.4\,
\left(\frac{\kappa_\mathrm{t}}{1-l}\right)^{1/3}\, 
\frac{\mu_{30}^{2/3}\,t_\mathrm{vis,100}^{1/3}}{\,P_{\star}^{2/3}\,M_{1.4}^{1/2}\,
M_\mathrm{disk,23}^{1/3}\,r_\mathrm{out,\odot}^{1/6}}\, .
$$
The ratio can be rather close to unity. The normalizations are 
$10^{23}$~g for the disk mass, $10^{30}$~[G\,cm$^3$] for the magnetic
dipole, 100~days for the viscous time, \rsun{} for the outer disk
radius. Situation of $r_\mathrm{mag} \gtrsim r_\mathrm{cor}$ is studied
by~\citet{dangelo-spruit2010}, who argue that no considerable expulsion
of matter from the disk is expected in such regime, as contrasted to the
propeller scenario. Using the magnetosphere radius as the inner disk
radius, we find the steady disk luminosity:
$$
L_\mathrm{dead} \approx 1.8\times 10^{35}
\,M_\mathrm{disk,23}^{3/2}\,M_{1.4}^{5/4}\, 
r_\mathrm{out,\odot}^{3/4} \, t_\mathrm{vis,100}^{-3/2}\,\mu_{30}^{-1}
\, \kappa_\mathrm{t}^{-1/2} \mbox{~~erg s}^{-1}\, 
$$
for $r_\mathrm{in} \ll r_\mathrm{out} $ from the both sides,
for typical $l\sim 1/3$ (see Table~\ref{tab:zeroes}). The upper limit on
the luminosity of the disk with conserved mass 
is estimated when the inner radius equals
the corotation radius:
$$
L_\mathrm{dead,max} \approx 4.4\times 10^{35} \,\mu_{30}^2\,
P_\star^{-3} \, m_{1.4}^{-1} \mbox{~~erg s}^{-1}\, .
$$
The regime can be sustained for sufficiently strong magnetic 
dipole and fast rotation of the neutron star:
$$
\mu_{30} P_\star^{-1} \gtrsim 0.63 \sqrt{(1-l)/\kappa_\mathrm{t}} \, 
m_{1.4}^{3/4}\, M^{1/2}_\mathrm{23}\,t^{-1/2}_\mathrm{vis,100}\,
 r_\mathrm{out,\odot}^{1/4} \, .
$$

\begin{figure}
\centering
\includegraphics[width=0.43\textwidth]{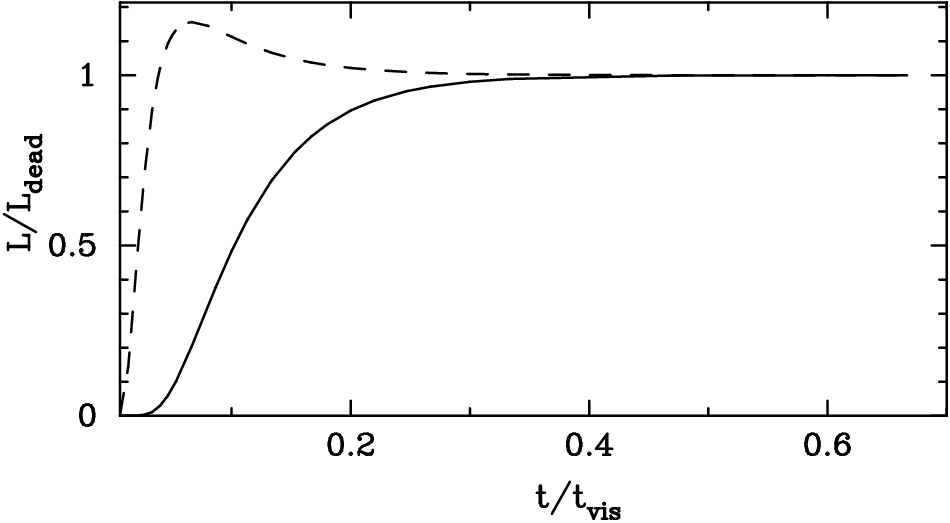}
\caption{Variation of the luminosity of the disk with zero accretion
rate through the inner radius normalized to value 
$L_\mathrm{dead}\equiv F_\mathrm{d}\,\omega_\mathrm{in}$.
 The disk starts from the outer radius 
(the solid line) or from $r=0.5^{4l} \times r_\mathrm{out}$ (the dashed
line), corresponding to the Green's functions in 
Fig.~\ref{fig.green_finiteB1}.
\label{fig.lumin_zeroAi}
}
\end{figure}

The flux of the angular momentum $F$ at small radii and the inner radius
of the disk always manage to adjust themselves to the outer variations
that proceed on longer viscous times. The steady value $F_\mathrm{d}
\propto \nu_0\,\rout^{b-3/2}$ could be low due to a degradation of
$\nu_0$ in the outer disk with low luminosity. Then the normalized
viscous time in the formulae above may be greater accordingly to
$t_\mathrm{vis} \propto \nu_0$.

Luminosity during the spreading of the ring 
(Fig.~\ref{fig.green_finiteB1}) is calculated by integrating expression
\eqref{eq.Qvis} over both surfaces of the disk confined between the
outer and inner radius; it is plotted in Fig.~\ref{fig.lumin_zeroAi}.
The inner radius of the disk is varying according to
Eq.~\eqref{eq.rmag_cond} and depends on the star's magnetic dipole. Its
typical value $r_\mathrm{mag,d}=r_\mathrm{cor} \approx 2.4 \times
10^{-3} P_{\star}^{2/3} \, M_{1.4}^{1/3} \, $\rsun.

The torque transfers the angular momentum of the central star to the
disk and, via the disk, to the orbital motion of the binary. 
The corotation radius increases gradually and, eventually, accretion on the center
begins~\citep{sunyaev-shakura77e,dangelo-spruit2011}.  
The characteristic brake time is the angular momentum of the star
 $2\,\pi\,I_{\star}/P_{\star}$ divided
by the torque $F_{\mathrm{d}}$:
$$
\frac{t_\mathrm{brake}}{t_\mathrm{vis}} \approx
 10^4  \, 
\frac
{I_{\star,45}}{M_{1.4}^{1/2}\,r_\mathrm{out,\odot}^{1/2}\,P_{\star}\,M_\mathrm{disk,23}}\,
,
$$
where $I_{\star,45} = I/10^{45}$~g\, cm$^2$ and a typical value of $l$ 
is substituted. If the mass of the disk grows due to the matter income,
the braking time of the star shortens, while the magnetosphere radius
decreases at the same time.

\subsection{The case of an arbitrary accretion rate at 
$r_\mathrm{out}$}

For a variable mass inflow $\dot
M_\mathrm{out}(t)$ at the outer disk edge $r_\mathrm{out}$ the
solution can be found by a procedure described in the Appendix B
\begin{equation}
\begin{split}
& F (x,t) = \int\limits_0^1 
\left(F_0(x_1)-\frac{x_1^{4l}}{2}\,h_\mathrm{out}\dot
M_\mathrm{out}(t)\right)\, G(x,x_1,t)\,
\mathrm{d} x_1~ +  \\
&+\frac{h_\mathrm{out}}{2}
\iint\limits_{0~~0}^{t~~1}   
\left[\frac{8}{x_1^2}\, \left(\frac{l}{\kappa}\right)^2\, 
\frac{ \dot M_\mathrm{out} (\tau)}{h_\mathrm{out}^{1/l}} -
\ddot M_\mathrm{out}(\tau)
\right]   x_1^{4l}\,
G(x,x_1,t-\tau)\, \mathrm{d} x_1\, \mathrm{d} \tau \,+\\
&~~~~~~~~~~~~~~~~~+ \frac{x^{4l}}{2}\,
\dot M_\mathrm{out}(t)\,h_\mathrm{out} \, ,
\end{split}
\label{eq:sol_nonzeroMdot_full}
\end{equation}
where $G(x,x_1,t)$ is given by (\ref{eq:green_zeroBi}) or
(\ref{eq:green_zeroAi}). Substituting $G$ by $G_{\dot M}$ in the
integrals above, and dividing the result by $h_\mathrm{out}$, we find
the accretion rate in the disk $\dot M(x,t)$. We have tried the formula
obtained in the numerous tests. In particular, it successfully
reproduces the results obtained by numerical simulations
of~\citet{minesh1994}. The further details, however, are out of the
scope of the present work.

\section{The accretion disk at the decay stage}
\label{s.decay}
We return now to the standard accretion disks with zero stress tensor at
the center.
For large $t$, the first term of the sum \eqref{eq:Mdot_green_zeroBi}
dominates, and a single exponential dependence on time prevails:
$$
G_{\dot M}(0,x_1,t)\Big|_{t> t_\mathrm{vis}} =
\frac{k_1^l\, x_1^{1-l}\,}{2\, l\,\Gamma(l)} \, 
\frac{J_l(k_1\,x_1)}{J_l^2(k_1)}\,
 \exp\left(-\frac{t\,k_1^2}{2\,l\,t_\mathrm{vis}}\right)~.
$$
 This corresponds to
the well-known exponential dependence on time of the accretion rate in
the disk with the time-independent viscosity $\nu$. The characteristic
time of the exponential decay is
\begin{equation}
t_\mathrm{exp} = h_\mathrm{out}^{1/l} \,\frac {\kappa^2} {k_1^2}  = 
\frac{16 \, l^2}{3\, k_1^2}\,
\frac{r_\mathrm{out}^{2}}{{\nu_\mathrm{out}}}~,
\label{eq:texp}
\end{equation}
where we use $\nu_\mathrm{out} = \nu_0 \, r_\mathrm{out}^b$. In
Table~\ref{tab:zeroes} we give locations $k_1$ of the first zeroes of
equation (\ref{eq:out_boundary}) for several values of $l$.
 The table also lists the 
numerical factors in the expressions for the rise and decay time,
\eqref{eq.t_max_inf} and \eqref{eq:texp}.

\begin{figure}
\centering
\includegraphics[width=0.45\textwidth]{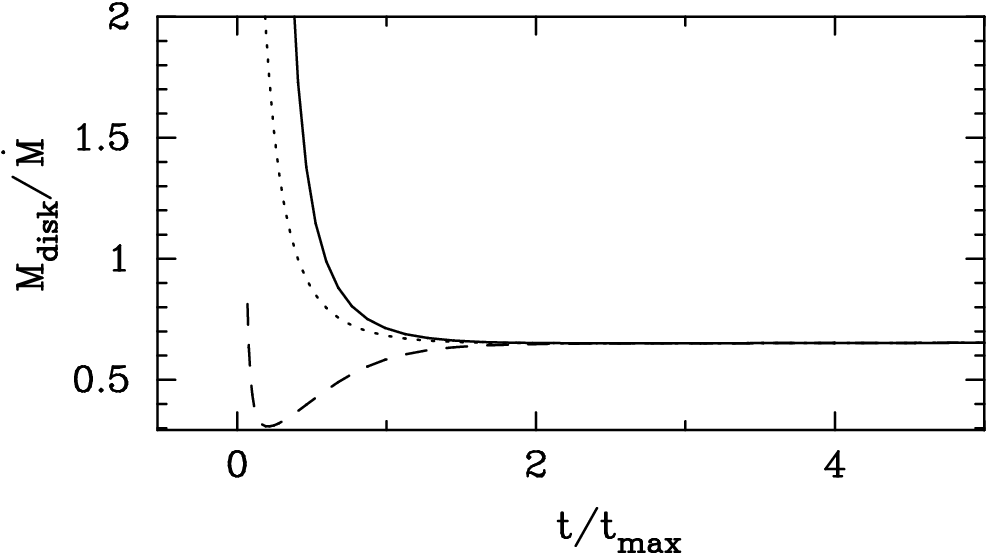} 
\caption
{Ratio of the disk mass to the inner accretion rate vs. time
normalized by the peak time. The
constant value indicates a self-similar distribution in the disk
has developed. The line styles
and parameters are as in Fig.~\ref{fig.mdotin}.
\label{fig.mdisk_mdot}
}
\end{figure}

At late times, the profile of $F(x,t)$ is self-similar, because
$G(x,x_1,t\rightarrow\infty) \rightarrow x^l\, J_l(k_1,x)$. The
distribution $\Sigma(r)$ is also self-similar. Actually, this profile
develops already at the very beginning of the exponential decay. This is
illustrated in Fig.~\ref{fig.mdisk_mdot} by the fact that the ratio of
the inner accretion rate to the disk mass, which is the integral of the
surface density, approaches a constant value that depends only on the
form of the radial distribution. In Table~\ref{tab:zeroes}, the values
of the parameter $a_0= \dot M_\mathrm{in} \, h_\mathrm{out} /
F_\mathrm{out}$ can be found. Parameter $a_0$ describes the self-similar
profile of $F(h)$ and was calculated by~\citet{lip-sha2000} for the
non-stationary finite $\alpha$-disks. For the sta\-tio\-nary disks with
non-zero accretion rate, $a_0=1$.

What does the presence of this profile mean for the released energy in
the disk? Using the energy balance equation for the optically thick
disks, we obtain that at the outer radius $ \sigma T_\mathrm{eff}^4 = 
(3/8\pi)\, \dot M_\mathrm{in}\, \omega^2/a_0$, i.e. the outer ring
luminosity is by factor $a_0\approx 1.44 $ less for the viscously
evolving disk ($l=2/5$) than in the case of the sta\-tio\-nary disk. The
factor $a_0(r)\approx 1.16$ for $r=0.5\times r_\mathrm{out}$.

\subsection{Comparing with the $\alpha$-disks}
\label{ss.alphadisks}

Let us now compare the finite-disk solution 
for the  time-independent $\nu$ and that obtained for the
turbulent viscosity in the general form $\nu = \nu_0\, \Sigma^a(t) \,
r^b$. For the latter form of viscosity, 
Eq.~\eqref{eq:basic_tda_Kepler} acquires the
following view:
\begin{equation}
\frac{\partial F}{\partial t}=
D\,\frac{F^m}{h^n}\,\frac{\partial^2F}{\partial h^2}~,
\label{eq.nonlin-diff}
\end{equation}
embracing a dimension constant
$$
D = \frac{a+1}{2}\, (G\,M)^2\, \left(\frac 32\,
\frac{\nu_0}{(2\,\pi)^a \,(G\,M)^b}\right)^{1/(a+1)}~,
$$
and dimensionless parameters $m$ and $n$:
$$
m=\frac{a}{a+1}\, , \qquad n = \frac{3\,a+2-2\,b}{a+1}\, .
$$
For a finite disk with such type of viscosity 
the inner accretion rate decays as a power law~\citep{lip-sha2000,dubus_et2001}:
\begin{equation}
\dot M (t) = \dot M (0) \, (1+t/t_0)^{-1/m}\, ,
\label{eq.power-law}
\end{equation}
where $\dot M (0)$ is the accretion rate at the time $t=0$,
which can be attributed to any moment of the decay stage.  
Eq.~\eqref{eq.nonlin-diff} describes the viscous evolution of the
$\alpha$-disks.

The characteristic
time of the solution $t_0 = h_\mathrm{out}^{n+2}/(\lambda\,m\, D\,
F^m_\mathrm{out}(0))$, where $\lambda$ is a numerical constant that can be
calculated for specific $a$ and $b$. Using the expression for $D$ and
relation $F_\mathrm{out} = 3\, \pi\, h_\mathrm{out}\,
\nu_\mathrm{out}\, \Sigma_\mathrm{out}$, we obtain
\begin{equation}
t_0 = \frac{4}{3\lambda\,a} \,
\frac{r_\mathrm{out}^{2}}{{\nu_\mathrm{out}(t=0)}}\, .
\label{eq.t_0}
\end{equation}
  For the
$\alpha$-disk in the Kramers' regime of the opacity, parameters $a=3/7$,
$b=15/14$, and  $\lambda=3.137$~\citep{lip-sha2000} should be used in 
Eqs.~\eqref{eq.nonlin-diff}--\eqref{eq.t_0}. 
In this case,  the dimensionless factor in~\eqref{eq.t_0} is
close to unity.  Value of $t_0$ depends on the
choice of the zero time.

\begin{deluxetable*}{ccccccl}
\tablecolumns{7} 
\tablewidth{0pt}  
\tablecaption{\label{tab:zeroes}
Parameters of the Green's functions for two types of inner boundary condition.
}
\tablehead{\colhead{$b$} & \colhead{$l$} &
\colhead{$k_1$} & \colhead{$\tmax^\infty (\rs^2/\nus)^{-1}$} & 
\colhead{$\texp (\rout^2/\nuout)^{-1}$} &
\colhead{$a_0$} & \colhead{Comment}
\\
\colhead{(1)} & \colhead{(2)} &\colhead{(3)} & \colhead{(4)} & 
\colhead{(5)}  & \colhead{(6)} & \colhead{(7)} 
} 
\startdata 
0     & $1/4$ & 1.0585 & $1/15$ & $ 0.298$ & 1.267 & constant $\nu$ \\
$1/2$ & $1/3$ & 1.2430 & $1/9$ &$ 0.383 $ &1.363 & $\alpha-$disks with constant $h/r$; $\beta-$disks \\
$3/5$ & $5/14$ & 1.2927  & $ 0.125$  & $ 0.407 $ &  1.392 & $\alpha-$disks, Thomson
opacity ($a_0=1.376^1$) \\
$3/4$ & $2/5$ &  1.3793     & $ 0.152$  &$ 0.449$  & 1.444& $\alpha-$disks, Kramers'
opacity  ($a_0 = 1.430^1$)\\  
1     & $1/2$ & 1.5708 & $2/9$ & $ 0.540$  & 1.571 & $F(h) \propto \sin
((\pi/2)\, h/h_\mathrm{out}  )$ \\
2      & $\infty$ &  ---  & ---  & --- & --- & $t_\mathrm{visc}$ does not depend on $r$ \\
\hline
\\
$2/5$ & $5/16$ & 3.4045 & $ 0.189 $ & ---& 0 & dead $\alpha-$disks, Thomson
opacity \\ 
$3/5$ & $5/14$ & 3.3425  &  $  0.265$  & --- & 0 &   dead $\alpha-$disks, Kramers' opacity
\enddata 
\tablecomments{
Columns are as follows: 
(1) The index of the radial dependence of $\nu$; 
(2) $l$ defined by \eqref{eq:consts_lb};
(3) The first zero of equations (\ref{eq:out_boundary}) and (\ref{eq:out_boundary_neg}), in the upper and lower part, respectively; 
(4) The dimensionless factor in \eqref{eq.t_max_inf} (the upper part)
and the one in a similar formula derived by LP74 for the time of maximum $F$ in the disk without central accretion~(the lower part of the Table); 
(5) The dimensionless factor in \eqref{eq:texp}; 
(6) Parameter describing the self-similar radial profile, $a_0= \dot M_\mathrm{in} \, h_\mathrm{out} / F_\mathrm{out}$;
(7) Note on the applicability. The opacity law is indicated for the $\alpha$-disks. The applicability of parameters is approximate for the $\alpha$-disk. 
Refs: $^1$\citet{lip-sha2000}.
}
\end{deluxetable*} 

In the Shakura--Sunyaev disks, the viscous stress is
proportional to the total pressure in the disk, and this proportionality
is parametrized by the $\alpha$-parameter~\citep{shakura1972}:
$$
w_{r\varphi} = \alpha \, \rho \, \vsound^2\, .
$$
Then the kinematic viscosity can be expressed using
(\ref{eq:stress_kepl_av})  as
\begin{equation}
\nut = \frac 23\, \alpha\, \frac{\vsound^2}{\omegak} \, 
\label{eq.nut_alpha}
\end{equation}
or
\begin{equation}
\nut =  \frac 23\, \alpha\, \omegak\, r^2\, 
\left(\frac{z_0}{r}\right)^2 
\, .
\label{eq.nut_alpha_z_r}
\end{equation}

To approximate the viscous evolution of the $\alpha$-disk by the disk
with steady viscosity, one has to estimate the appropriate parameter $b$
for Eq.~\eqref{eq:nut} and \eqref{eq:lin-diff}. This can be effectively
done using the relation $\nut \propto r^{1/2}\, (z_0/r)^2 \propto r^b$.
The solution for the stationary pressure-dominated stationary
regions~\citep{sha-sun1973} gives the relative half-thickness $z_0/r
\propto r^{1/8}$, consequently, $b\simeq 3/4$ or $l\simeq 2/5$ in
\eqref{eq:lin-diff}, for the free-free opacity regime.
Table~\ref{tab:zeroes} lists the parameters approximating in a similar
way the standard $\alpha$-disk with the Thomson opacity and the `dead'
disks models~\citep{sunyaev-shakura77e}.

\begin{figure}
\centering
\includegraphics[angle=270,width=0.45\textwidth]{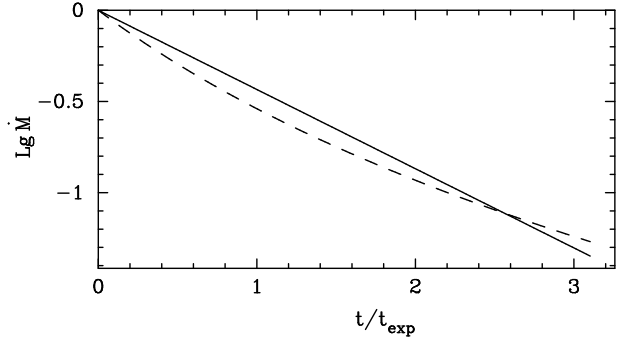}
\caption{
Relative variation of the accretion rate in two cases: $\nu = \nu_0\, 
\, r^b$ ($b=3/4$, the solid line) and $\nu = \nu_0\, \Sigma^a
\, r^b$ (Kramers' opacity, $a=3/7$, $b=15/14$, the dashed line). 
Disk parameters $r_\mathrm{out}$ and $\nu_\mathrm{out}$ are the same in
the both models.
\label{fig.teordep}
}
\end{figure}

In Fig.~\ref{fig.teordep} the two laws are plotted: exponential $
\propto \exp (-t/t_\mathrm{exp})$ for $l=2/5$ and power-law
\eqref{eq.power-law} for $m=3/10$. Viscosity at the outer radius and the
outer radius itself are the same. If applied to the real bursts during
the time spans of order of $\sim t_\mathrm{exp}$, the models cannot be
easily discriminated~\citep[this was also pointed at
by][]{dubus_et2001}. This fact provides a possibility to estimate
$\alpha$ using the $e$-folding time of the light curve. In the next
Section, we show how it could be done using the observations of the
disks in X-ray novae, whose light curves in many cases show exponential
decays.

\section{The burst light curves of X-ray novae}
\label{s.xnovae}

In the X-ray novae, binary systems with a compact star and, typically, a
less-than-solar-mass optical star, transient outbursts are thought to
originate due to a dwarf nova-type
instability~\citep[e.g.,][]{lasota2001}. The temperature rises in a
portion of the `cold' neutral disk, which becomes ionized, and the
viscosity $\nu$ rises. Moreover, the turbulent parameter $\alpha$
changes from $\alpha_\mathrm{cold}$ to higher $\alpha_\mathrm{hot}$. The
heating front moves outward from some ignition radius with the speed
$V_\mathrm{front} \sim \alpha_\mathrm{hot}\, \vsound$. It stops at a
radius, where the surface density of the disk is too low to sustain the
stable (on the thermal time-scale) hot
state~\citep{meyer1984,menou_etal2000,dubus_et2001,lasota2001}. 

In the dwarf novae, the hot state accretion is quenched swiftly
(comparing to the viscous time) by the matter returning to the cold
state. It was suggested that in the X-ray novae the illumination from
the center has a stabilizing effect on the outer disk, keeping the 
temperatures above the hydrogen ionization temperature
\citep{meyer-meyer1984,chen_et1993}. \citet{kin-rit1998} propose that in
order for the disk to undergo the viscous evolution an intensive
irradiation must be included.

 A canonical light curve of an X-ray nova is described as a 
FRED~\citep{chen_et1997}. Theoretically, if the disk is truncated from
outside, a steeply decaying light curve is
produced~\citep{kin-rit1998,lip-sha2000,wood_etal2001}. The main
property is that the disk does not expand infinitely, which is a
consequence of its being in a binary system, and its radius does not
change on a viscous time scale. Scenario by \cite{kin-rit1998} produces
exponential decays due to the steadiness of the parameter $\nu$ in time.
More realistic temporal dependence of $\nu$ leads to a steep power-law
decay~\citep{lip-sha2000,dubus_et2001}, which is close to an exponential
one during few viscous times (see the previous Section). 

The possibility to model the X-ray novae bursts, which have the FRED
profile, by the viscously evolving disk with the steady viscosity and
after a descrete mass-transfer event, was shown by~\cite{wood_etal2001}
(see also the modeling in~\citealt{Sturner-Shrader2005}). If a burst is
due to an instability affecting a slab of matter extended along the
radii, the Green's function obtained in the present work should be used.
Effectively, after a heating front has passed, the dwarf-nova type
instability leads to some distribution of the hot-ionized matter with
high viscosity. In the most trivial case, when the bulk mass is
concentrated near $\rout$, the evolution of the disk approaches the one
found by \cite{wood_etal2001}.

Let us compare the typical time of the heating front propagation
$r/V_\mathrm{front}$ and the time $t_\mathrm{max}$ it takes the inner
accretion rate to peak, which is of order of 
$t_\mathrm{max}^\infty$~(see \eqref{eq.t_max_inf} and
Fig.~\ref{fig.Cmax1}). Relating $\alpha$ and $\nu$ by 
\eqref{eq.nut_alpha}, we obtain:
$$
\frac{t_\mathrm{front}}{t_\mathrm{max}} =  \frac{l+1}{2\,l^2} \,
\sqrt{\frac{r_\mathrm{s}\,\vsound^2 }{ G\, M}}~ \sim 0.1 \sqrt
{\frac{T_4\, r_{11}}{\mu\, m_x}}\, ,
$$
where the gas temperature is normalized by $10^4$~K, central star mass
by a solar mass, and radius by $10^{11}$~cm, $\mu$ is the molecular
weight of the hot matter. Radius $r_s$ is the characteristic radius
where the bulk mass resides at the beginning of an outburst. This ratio
indicates that we are safe to model an X-ray nova outburst at time $\sim
\tmax$ using the Green's function obtained.

If one assumes the commonly suggested surface density distribution in
the quiescent state in the X-ray novae, $\Sigma \propto
r$~\citep{lasota2001}, then, as it is shown in Fig.~\ref{fig.mdotin},
after the peak of the burst, the only important parameters are the disk
mass and the outer radius. For such  density distribution, the bulk
mass of the disk resides  close to the outer radius. As we show in
Sect.~\ref{s.decay}, already at the maximum of an outburst the disk
acquires the self-similar radial distribution, which is not dependent on
the initial distribution of matter. The necessary condition for this is
that the bulk mass of the viscously evolving disk is located near the
outer radius. This is consistent with the absence of a power-law
interval before the exponential decline on the the light curves of the
X-ray novae that show the FRED behaviour.

\begin{figure*}
\centering
\includegraphics[angle=270,width=0.32\textwidth]{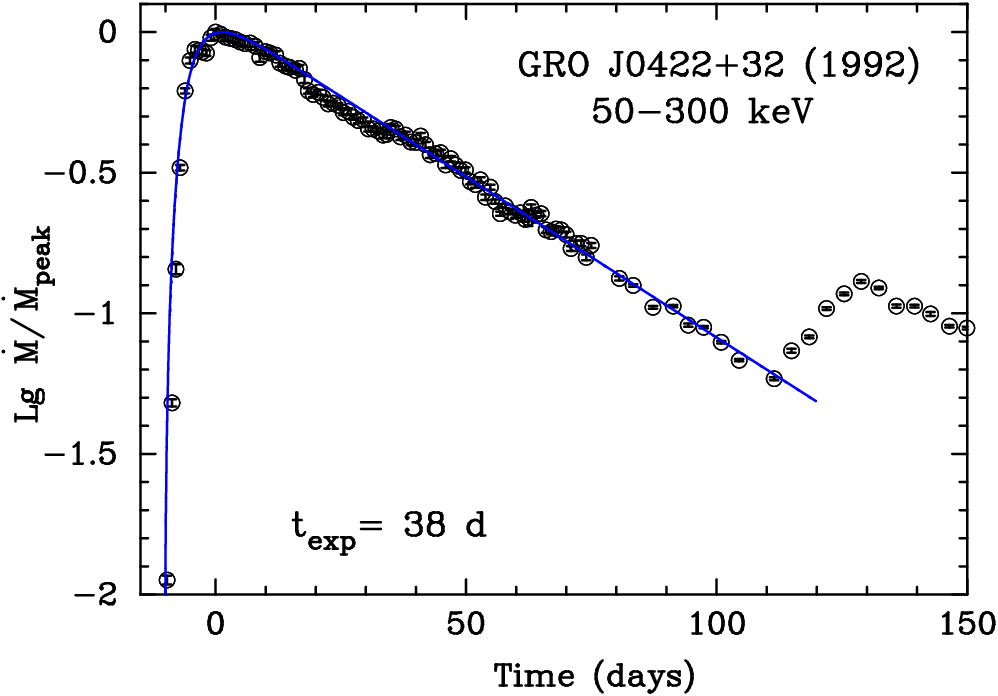}
\includegraphics[angle=270,width=0.32\textwidth]{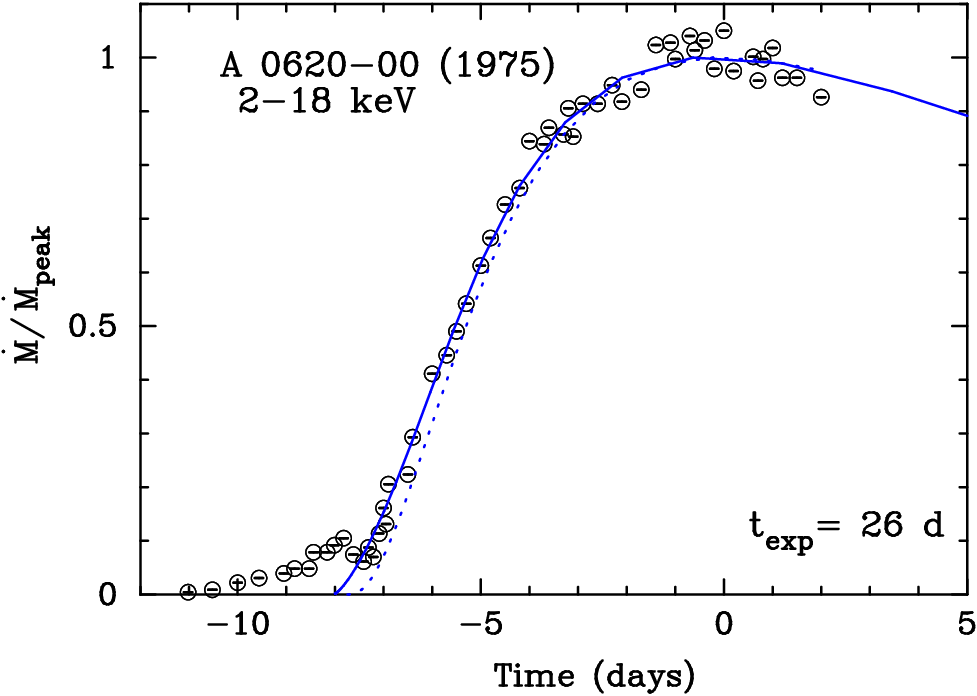}
\includegraphics[angle=270,width=0.32\textwidth]{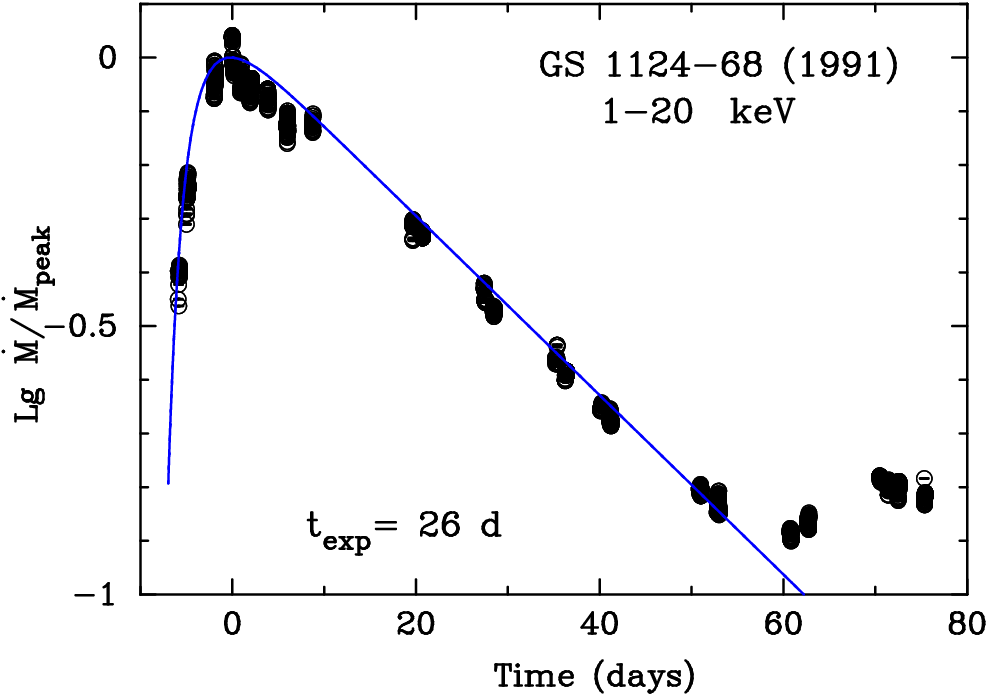}
\includegraphics[angle=270,width=0.32\textwidth]{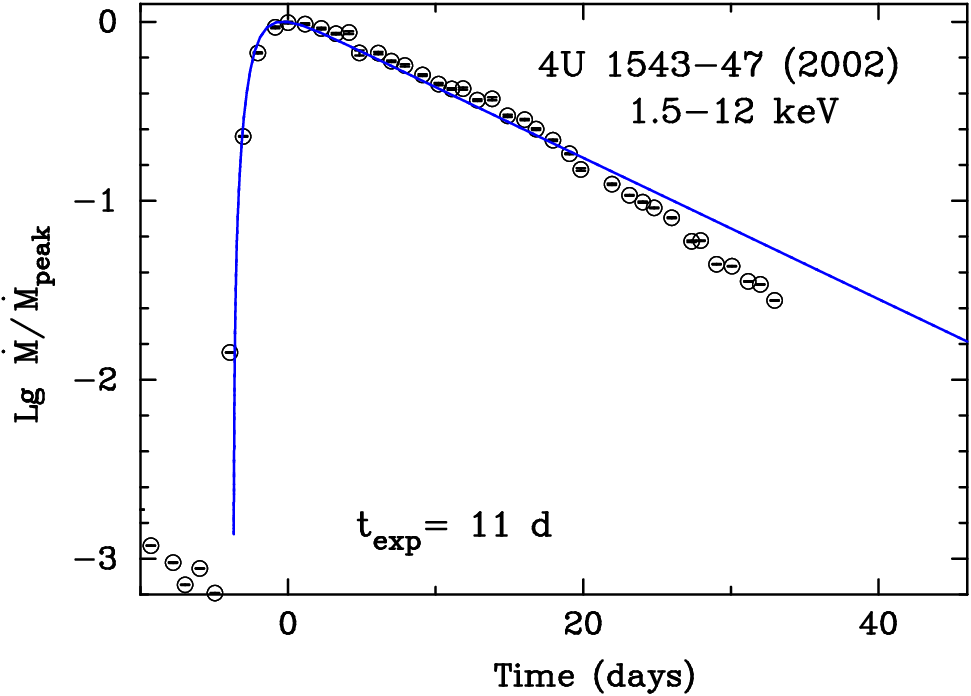}
\includegraphics[angle=270,width=0.32\textwidth]{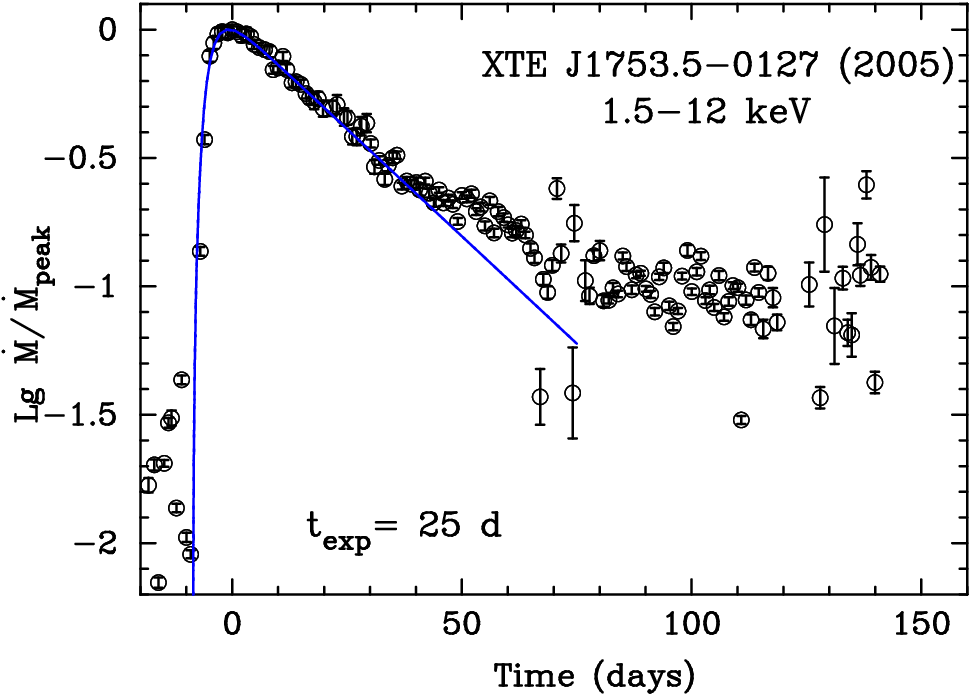}
\includegraphics[angle=270,width=0.32\textwidth]{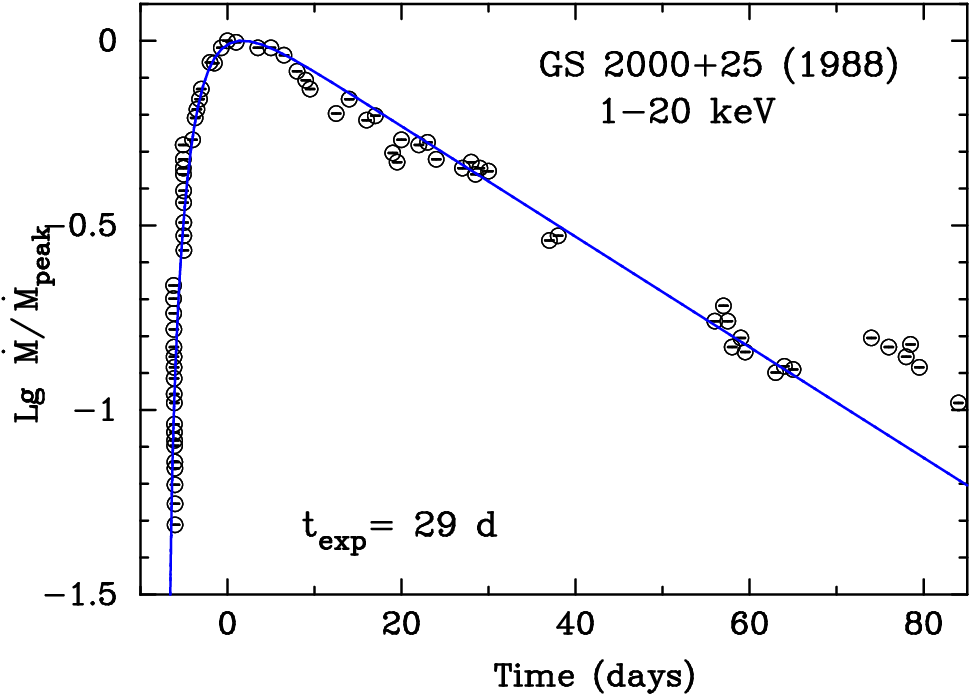}
\caption{
The peak-normalized light curves of GRO\,J0422+32 in 1992, A\,0620-00 in
1975, GS\,1124-68 in 1991, GS\,2000+25 in 1998 (as collected in
\citealt{chen_et1997}) and 4U\,1543-47 in 2002 and XTE J1753.5-0127 in
2005 (the quick-look results provided by the ASM/RXTE team). The energy
bands of the X-ray data are indicated in the plots. The model curves are
the peak-normalized accretion rate, calculated by
\eqref{eq:mdot_from_green} using $\texp$ indicated for each burst, and
$l=2/5$. The initial surface density distribution $\Sigma\propto r$,
and the inner radius of the initial hot zone is $0.01\times\rout$. 
Note the different axis scale for A\,0620-00,  for which
two different models are plotted, with the initial inner radius at
$0.001\times \rout$ (solid line) and $0.3\times \rout$ (dotted line).
\label{fig.srcs_lc}
}
\end{figure*}

\begin{figure}
\centering
\includegraphics[width=0.45\textwidth]{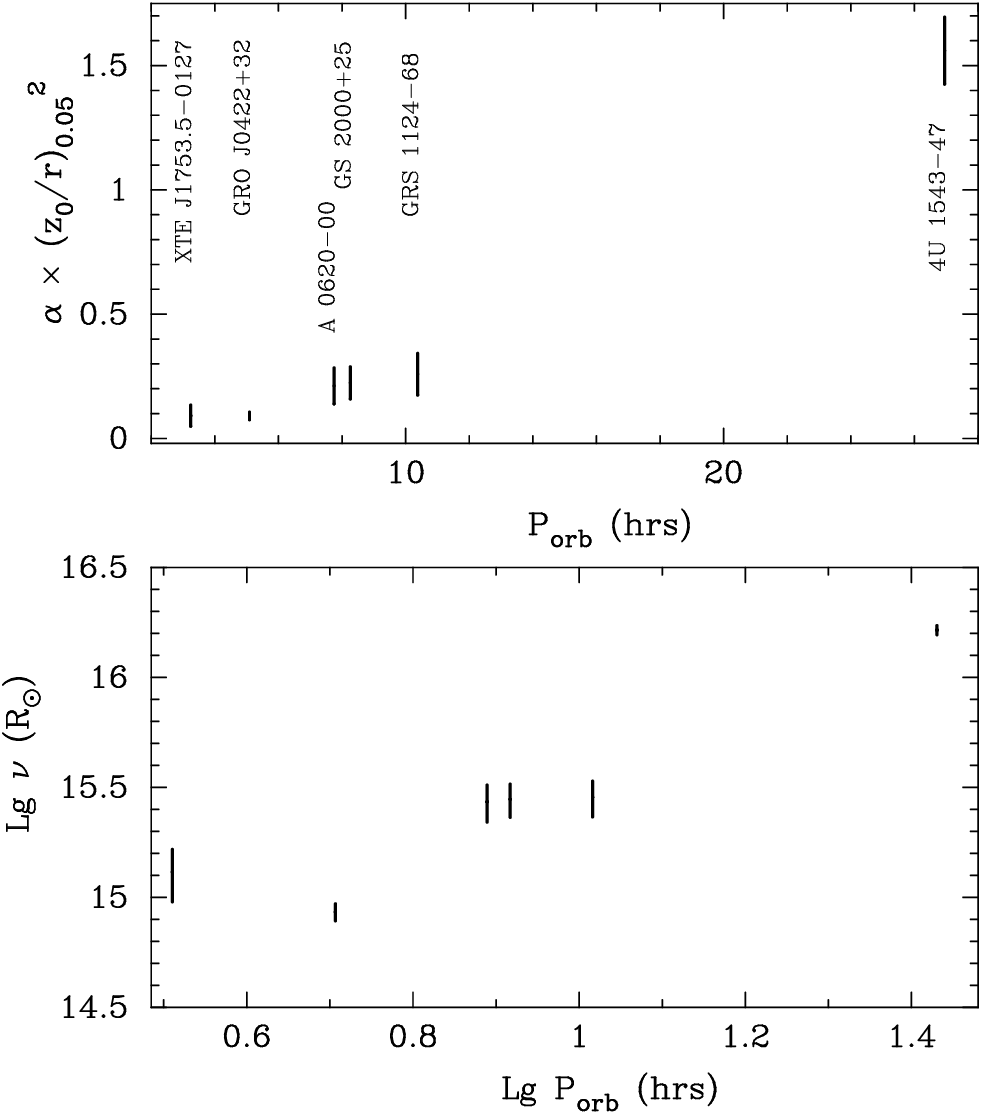}
\caption{
Top panel: estimates for the turbulent parameter $\alpha$ times
the square relative half-thickness of the disk at $\rout$ for the six bursts
of X-ray novae. 
Lower panel: corresponding estimates of the viscosity parameter $\nu$ at $r=$\rsun.
\label{fig.alpha_P}
}
\end{figure}

Eqs.~\eqref{eq.t_max_inf} and \eqref{eq:texp} yield the ratio
$t_\mathrm{exp}/t_\mathrm{max}$ taking into account that
$t_\mathrm{max}$ is very close to $t^\infty _\mathrm{max}$ for
$r_\mathrm{s} \sim r_\mathrm{out}$. The value of
$t_\mathrm{max}/t_\mathrm{max}^\infty$ is depicted in
Fig.~\ref{fig.Cmax1}, where $t_\mathrm{max}$ is the peak time for the
finite disk. For $l=2/5$ (see Table~\ref{tab:zeroes}) we obtain
$t_\mathrm{exp}/t_\mathrm{max} \approx 3$. This provides a rough test
whether the particular light curve is produced by the viscously evolving
disk made of hot ionized matter and with  approximately constant
outer radius. 

The viscous-disk relation $t_\mathrm{rise} \approx 0.15\,
\rout^2/\nuout$ for the hot-disks with the Kramers' opacity in the outer
parts leads to a conclusion that to obtain a secondary maximum on the
X-ray light curve of an X-ray nova, the disk has to provide an extra
mass input at the radius that is determined by the time of the input.
The secondary maximuma are usually observed around $(1-2)\, \texp$ after
the peak. If the extra mass input happens at the time of the burst
maximum then it should happen at least at a radius about $2.4 \, \rout$
(from $t_\mathrm{exp}/t_\mathrm{rise} \sim 3$ and $\nu \propto
r^{3/4}$), without mass input from the intermediate radii. Another
possibility is that the secondary maximum is triggered later after the
peak and then it can happen at a radius $\sim \rout$ (as in the model of
\citealt{ertan-alpar2002}).

If an X-ray nova has the viscously evolving accretion disk with constant
$\rout$ during the interval of the `exponential decay' (usually, first
$\sim 50$~days after the peak), we can infer the value of the turbulent
parameter $\alpha$, using the closeness of the model with steady
visosity and the $\alpha$-disk. As the irradiation, which
provides the steadiness of the hot zone outer radius, does not 
change the vertical structure of the disk~\citep{dubus_et1999}, we can
adopt $l=2/5$ applicable for the hot ionized standard $\alpha$-disks.
Combining Eqs.~\eqref{eq:texp} and \eqref{eq.nut_alpha_z_r}, we get
\begin{equation}  
\alpha \approx 0.15
\,\left(\frac{r_\mathrm{out}}{2\,R_{\sun}}\right)^{3/2}
\left(\frac{z_0/r_\mathrm{out}}{0.05}\right)^{-2}  \left(\frac{M}{10\,
M_{\sun}}\right)^{-1/2} 
\left(\frac{t_\mathrm{exp}}{30^\mathrm{d}}\right)^{-1} ~,
\label{eq.alpha_est} 
\end{equation} 
which relates the $\alpha$-parameter
to the observed  $t_\mathrm{exp}$, the \hbox{$e$-folding} time of the bolometric
luminosity or the inner accretion rate. The last formula resembles quite
a few others found in  the literature. Similarly, the relation of
$\alpha$ to the half-thickness of the disk cannot be eluded.  However,
an advatage is that \eqref{eq.alpha_est} contains  the e-folding time
which can be inferred quite accurately from observations.

We have sampled six FRED light curves in different X-ray novae
(Fig.~\ref{fig.srcs_lc}) and fitted them with the steady-$\nu$
viscously-evolving disk model. For particular values of $\texp$, the
model curves agree well with the peak-normalized count-rates before the
second maximum, which is observed in most of the cases.  For
A\,0620-00, we use the value of $t_\mathrm{exp}$, estimated by
\citet{chen_et1997} from the $3-8$~keV light curve during the decline.
Variation of the initial inner radius of the hot zone leads to slight
changes of the model curve for the particular initial distribution of
$\Sigma(r)$, as can be seen in the panel for \hbox{A\,0620-00}.

In Fig.~\ref{fig.alpha_P} we plot the estimates for $\alpha$-parameter
times the  normalized square of the relative half-thikness at the
outer radius of the disk.  Theses estimates are obtained using
\eqref{eq.alpha_est} with the values $\texp$ indicated in
Fig.~\ref{fig.srcs_lc}. In the lower panel, the viscosity $\nu$ is
plotted, found accordingly to \eqref{eq:texp}. As $\nu$ is not constant
over radius, we choose to present its value at the same radius \rsun{}
for all cases. The considered X-ray novae have known orbital periods.
The masses of the primary and secondary components, mass ratios are from
\citet{Charles-Coe2006}. For black hole candidate XTE\,J1753.5-0127 with
$P_\mathrm{orb}=3.24$~hr~\citep{zurita_etal2008} we set rather
arbitrarily $M=3-15$~\msun{} and $M_\mathrm{sec}=0.2-1.2$~\msun. The
Roche lobes $\rrl$ are calculated using the formula
by~\citet{egglton1983}, and $\rout/\rrl = 0.88\pm 0.02$. The vertical
barrs in Fig.~\ref{fig.alpha_P} reflect the errors related to the
uncertainties of the binary parameters and do not take into account
uncertainty of $\texp$. 

\citet{suleimanov_et2007e}  give the half-thickness of the outer
parts for the stationary $\alpha$-disks when the hydrogen
is completely ionized: 
$$z_0/r \approx 0.05 \,(M/\mbox{\msun})^{-3/8}\,
\dot M_{17}^{3/20} \, r_{10}^{1/8}\, \alpha^{-1/10}\,  ,$$ with the parameters'
normalizations equal to $10^{17}$\,g\,s$^{-1}$ and $10^{10}$~cm. It can be
rewritten for the factor in  Fig.~\ref{fig.alpha_P} as follows
$$
\left(\frac{z_0/r} {0.05} \right)^2 \approx 
1 \times \left(\frac{M}{10\,\mbox{\msun}}\right)^{-9/20}\,
\left(\frac{\dot M}{0.1\,\dot M_\mathrm{cr}}\right)^{3/10} \, 
\left(\frac{\rout}{\mbox{\rsun}}\right)^{1/4}\, 
\alpha_{0.1}^{-1/5}\,,
$$
where $\alpha$ is normalized to 0.1 and $ \dot M_\mathrm{cr}$ is the
conventional rate of accretion producing the critical luminosity: $
0.1\, \dot M_\mathrm{cr}\, c^2 = 1.3 \times 10^{38}
(M/\mbox{\msun})$~erg\,s$^{-1}$. For the black hole masses in the range
$3-15$~\msun{}, accretion rates in $(0.01-0.5)\, \dot M_\mathrm{cr}$,
$\alpha$ in $0.1-0.5$, and $\rout$ in $1-3$\,\rsun, the above factor
changes between extremes $\sim 0.3$ and $\sim 3.8$. For 4U\,1543-47 with
parameters $\dot M/\dot M_\mathrm{cr} \approx 0.25$~\citep{wu_etal2010},
$M=9.4$~\msun~\citep{Charles-Coe2006},  $\rout = 4.5$~\rsun~$=0.9 \times
R_\mathrm{RL}$ and $\alpha=0.2$ one obtains the factor $(z_0/r/0.05)^2
\approx 1.8$. It follows from Fig.~\ref{fig.alpha_P} that $\alpha$ can
be less than 1  in this case, but this is still not a
self-consistent value. Apparently, the disk in this longer-period X-ray
nova was not entierly involved in the viscous evolution that produced
the FRED light curve in 2002. Levels of $\alpha$ of the other five
bursts (Fig.~\ref{fig.alpha_P}, the upper panel) agree with the
estimates for the fully ionized, time-dependent accretion disks obtained
by various approaches in X-ray and dwarf
novae~\citep{King_et2007,suleimanov_etal2008,kotko-lasota2012}, in the
decretion disks of Be stars~\citep{carciofi_etal2012}.

The values of $\texp$, indicated in Fig.~\ref{fig.srcs_lc}, are to be
considered with caution. Note that the observed light curves are limited
to an energy band. Spectral modeling might be required in many cases to 
obtain accurate $\texp$. If an X-ray nova during the burst is in the
{\em high/soft} state, when the disk emission dominates the spectrum,
the spectral modeling can provide $T_\mathrm{in}(t) \propto \dot M^{1/4}
_\mathrm{in} (t)$ in a straightforward way. We expect $\texp$ not to
alter by a factor larger than 2 for the X-ray flux dominated by the disk
(preliminary resluts for 2-10~keV).

On the other hand, some X-ray novae bursts proceed entirely in the {\em
low/hard state}, such as the burst of
XTE\,J1753.5.-0127~\citep{Miller-Rykoff2007}. Still, the form of the
corresponding X-ray light curves is informatve. Even if the optically
thick disk does not reach the innermost orbits, it is very probable that
the radiated energy is proportional to the accretion rate, because the
released potential gra\-vi\-ta\-ti\-onal energy is the ultimate source
to power the source. Second, the innermost disk peculiarities do not
influence the behavior of $\dot M(t)$ as the latter is defined by the
process of the viscous matter redistribution in the outer disk.

\section{Discussion}

Equation \eqref{eq:basic_tda_Kepler} is written without the tidal-stress
term, following the proposition that it should be negligible everywhere
in the disk, except in a thin ring near the tidal truncation
radius~\citep{ich-osa1994}. According to \citet{pringle1991}, effects
near the tidal-torque radius can be approximated with an effective
boundary condition $\dot M =0$. Numerical models confirm that most of
the tidal torque is applied in a narrow region at the edge of the disk,
where perturbations become non-linear and strong spiral shocks
appear~\citep[][and references
therein]{pringle1991,ich-osa1994,hameury-lasota2005}. As
\citet{smak2000} notes, in the case of dwarf novae, during outbursts,
the observed values of the outer radius appear to be approximately
consistent with the theoretical predictions. \cite{hameury-lasota2005}
argue that the action of the tidal torques are important also well
inside the tidal radius. They notice at the same time that observations
do not allow discriminating between the rapidly growing and smoother
tidal stresses, as the model light curves are not strongly affected.

The tidal radius for all mass ratios can be approximated as $0.88\pm
0.02$ of the average Roche lobe radius of a given component in circular
binaries ~\citep{pap-pri1977}. The disk can extend beyond this radius
due to the high inclination of the binary orbit~\citep{okazaki2007b}.
Effects of the high eccentricity of the binary orbit on the disk
truncation radius have been studied by various methods
\citep[e.g.,][]{artymowicz-lubow1994,pichardo_etal2005,okazaki2007b}.

The gravitational influence of the secondary on the disk in a binary
system leads to a number of interactions, non-resonant (or tidal) and
resonant, any of those transferring angular momentum from the disk to
the binary~\citep{artymowicz-lubow1994}. The tidal distortions generally
lead to the truncation of the disk, whereas the resonances of different
strength impact or do not the disk structure, depending on their growth
rate and other conditions~\citep{whitehurst-king1991, ogilvie2002}. An
example of a resonance action is the debated explanation of the
superhamps in SU UMa stars~\citep[see,
e.g.][]{lubow1991a,kornet-rozyczka2000, hameury-lasota2005}.

In our mathematical set-up, we have presumed a fixed outer radius of
the disk without going into details of its actual value. Thus the model
is appplicable to the coplanar disks in low-eccentricity binaries. The
outer radius of the disk can also change due to intrinsic processes. 
Reproducing the whole evolutionary cycles of the dwarf novae and X-ray
novae, \citet{smak2000}, \citet{dubus_et2001} should have taken into account
the variations of the outer disk radius, but these changes are
significant during the time spans much greater than the viscous times. 

\section{Summary}

 Under some astrophysical conditions, viscously evolving disks 
formed in binary systems are effectively truncated from
outside. In the present work, we use the method proposed by
\citet{lust1952} to find the Green's functions for the linear viscous
evolution equation for the disk with the finite outer radius. Green's
functions are obtained for the viscous angular momentum flux $F$ and the
accretion rate $\dot M$. Two inner boundary conditions at the zero 
coordinate are considered. They correspond to the accreting disks and to
the disks with no mass sink at the center.
The Green's functions allow one to compute $\Sigma(r,t)$ and $\dot
M(r,t)$ for arbitrary initial surface density profiles that develop
into self-similar distributions on sufficiently long time scale. The
analytic formula to calculate the disk evolution with the variable outer
mass inflow is derived.
A solution,  which can be found with the use of the Green's
function, has the properties defined by the main equation of the viscous
evolution and thus provides the basic description of the transient
phenomena in the viscous disk.
 
The Green's functions are found in the form of  quickly converging
series and can be easi\-ly reproduced with standard computer methods. 
The C-code written for the case of the accreting disc with the use of
the GNU Scientific
Library\footnote{\href{http://www.gnu.org/software/gsl/}{http://www.gnu.org/software/gsl/}}
can be 
downloaded\footnote{\href{http://xray.sai.msu.ru/~galja/lindisk.tgz}{http://xray.sai.msu.ru/$\sim$galja/lindisk.tgz}}.

We present the relations between the rising time, $e$-folding time, and
disk viscosity. For six bursts in the X-ray novae, which are of FRED
type, we show that the model describes well the peak-normalized light
curves before the second maximum. This favors the mainly viscous nature
of their evolution during this period and enables us to  obtain
an estimate  of viscosity $\nu$, which depends on the outer disk radius. 

The models with time-independent viscosity are shown to approximate 
well the evolution of $\alpha$-disks during time intervals comparable to
the viscous time. Consequently, estimates of the $\alpha$ parameter may
be derived for the disks in the binaries with known period and masses.
These estimates rely on the relative half-thickness of the disk at the
outer radius. The estimates of $\alpha$ for the five short-period X-ray
novae outbursts, GRO\,J0422+32 in 1992, A\,0620-00 in 1975, GS\,1124-68
in 1991, GS\,2000+25 in 1998, and XTE J1753.5-0127 in 2005, are in line
with the values estimated so far for the hot viscous 
disks~\citep{King_et2007,suleimanov_etal2008,kotko-lasota2012}.

Another Green's fucntion is found for the disk that has zero accretion
rate at the inner radius and acquires angular momentum from the
central star. In the steady state, the mass of the disk cannot be very
high and the disk has low luminosity. It radiates most of the rotational
power transferred from the central star. This disk works as a gear
transmitting the angular momentum of the central star to the orbital
motion. The `dead' stage is ended by an abrupt fall or trickle of the
matter on to the star after the centrifugal barrier has moved close
enough or beyond the magnetosphere's radius.

\section*{Acknowledgements}

The author is grateful to N.~I.~Shakura, K.~Stempak, J.~L.~Varona,
M.~Perez and the anonymous referee for useful comments. The work is
supported by the Russian Science Foundation (grant 14-12-00146). 

\appendix

\section{Infinite disks: Green's function and basic relations}

Green's function for the infinite disk, obtained by LP74, can be written
in the dimensionless form with our designations (c.f. formula
\eqref{eq:green_zeroBi}) as follows:
\begin{equation*}
 G^\infty(x,x_1, t) = \frac{\kappa^2\,h_\mathrm{c}^{1/l}\, x^l\,
x_1^{1-l}}{2\,t}  
 \exp\left(-\frac{x_1^2+x^2}{4t}\,\kappa^2\, h_\mathrm{c}^{1/l}
\right) 
 I_l \left(\frac{x\,x_1}{2t}\, \kappa^2\, h_\mathrm{c}^{1/l}\right)~,
\end{equation*}
where $I_l$ is the modified Bessel function of the first kind, $x =
(h/h_\mathrm{c})^{1/2l}$, $h_\mathrm{c}$ is some characteristic value of
the specific angular momentum, which we use instead of $h_\mathrm{out}$.

To find the physical distribution of $F(x,t)$ in the course of the
evolution of the initial narrow ring of matter, initially located at the
coordinate $x_s$, one takes the integral of $F(x_1,0)$ (\ref{eq:Fring})
with the kernel $G^\infty$ on the interval $x_1\in[0,\infty]$:
\begin{equation*}
 F(x, t) = \frac{M_\mathrm{disk} \,h_\mathrm{c}\,l\,(x\,x_s)^l}{t} \,  
\exp\left(-\frac{x_s^2+x^2}{4t}\,\kappa^2\, h_\mathrm{c}^{1/l}
\right) 
 I_l \left(\frac{x\,x_s}{2t}\, \kappa^2\, h_\mathrm{c}^{1/l}\right)~,
\end{equation*}
where $M_\mathrm{disk}$ is the initial mass of the ring.   

The accretion rate at the center can be found from $\dot M_\mathrm{in}
= \partial F/\partial h|_{h\rightarrow 0} $ (see
(\ref{eq:mdot_from_green})). One finds 
$$
\dot M_\mathrm{in} (t) = \frac{x^{1-2l}}{2\,l\, h_\mathrm{c}} \,
\frac{\partial F}{\partial x}\Big|_{x\rightarrow 0} = \frac{M_\mathrm{disk} \tau_e^l} {\Gamma(l)}\, 
 \frac{e^{-\tau_e/t}}{t^{1+l}}\, ,
$$
or
$$
\dot M_\mathrm{in} (t) = \dot M_{\mathrm{in,\,peak}}^\infty  \left(
\frac{\tau_\mathrm{pl}}{t}\right)^{1+l}\, e^{-\tau_e/t}\, ,
$$
where the  typical times $\tau_\mathrm{pl}$ and $\tau_e$ are introduced,
$$
\tau_e = \frac{\kappa^2\, h_\mathrm{s}^{1/l}}{4} =  \frac{1+l}{e}\,
\tau_\mathrm{pl}\,  .
$$
We find that the accretion rate peaks at the time \eqref{eq.t_max_inf} $t_\mathrm{max}^\infty =\tau_e/(1+l)$ 
with the value
\begin{equation}
\dot M_{\mathrm{in,\,peak}}^\infty =  \frac{M_\mathrm{disk}}{t_\mathrm{max}^\infty} \, 
\frac{(1+l)^l}{e^{1+l}\, \Gamma(l)}\, .
\label{eq:ap.dotmmax_snfinite}
\end{equation}

\section{Solution for the case with variable accretion
rate at the outer edge of the disk}
For infinite disks, see 
\citet{metzger_etal2012} and \citet{shen-matzner2014}, who have obtained the
Green's function solution to the viscous evolution of a disk with mass
sources/sinks, which are distributed in a fashion along the disk
radius.
  
 We have the following problem with the inhomogeneous boundary and
initial conditions: 
\begin{equation}
\begin{split}
&\frac{\partial^2 F}{\partial h^2} = \frac 14\,
\left(\frac{\kappa}{l}\right)^2\, h^{1/l-2}\,
 \frac{\partial F}{\partial t}\, \\
& \partial F/\partial h|_{h=h_\mathrm{out}} = \dot M_\mathrm{out}(t) \\
&F\,|_{t=0} = F_0(h)\, .
\end{split}
\label{eq:problim_inh1}
\end{equation}
Let us substitute function $F(h)$ by a function $\tilde F(h)$ using the
relation
\begin{equation}
F(h) = \tilde F(h) + \frac{h^2}{2\,h_\mathrm{out}}\, 
\dot M_\mathrm{out}(t) \, .
\label{eq:substit}
\end{equation}
This substitution is used, for example, to solve the problem of finding
the temperature distribution in a cylinder, on a surface of which a
thermal flux is defined \citep{bokra1998er}. For
the new $\tilde F(h)$ the problem is the following:
\begin{equation}
\begin{split}
&\frac{\partial \tilde F}{\partial t}  = 4\,
\left(\frac{l}{\kappa}\right)^2\, h^{2-1/l}\,
\frac{\partial^2 \tilde F}{\partial h^2}  + \Phi(h,t)
 \, , \\
& \partial \tilde F/\partial h|_{h=h_\mathrm{out}} = 0\, ,\\
&\tilde F\,|_{t=0} \equiv \tilde F_0(h) = F_0(h) 
 -  \frac{h^2}{2\,h_\mathrm{out}}\, \dot M_\mathrm{out}(t=0)
 \, ,
\end{split}
\label{eq:problim_inh2}
\end{equation}
where 
$$\Phi(h,t) \equiv 4\, \left(\frac{l}{\kappa}\right)^2\, 
\frac{h^{2-1/l}\, \dot M_\mathrm{out}}{h_\mathrm{out}} -
\frac{h^2}{2\,h_\mathrm{out}}\, \frac{\mathrm{d} \dot
M_\mathrm{out}}{\mathrm{d} t}~.
$$ 
The corresponding Sturm--Liouville problem with a free variable
$\xi=h/h_\mathrm{out}$ is investigated in Sect.~\ref{sec:ev_fin_disk}:
it consists of Eq. (\ref{eq:sturm-liuv}) 
\begin{equation}
\frac{\partial^2f_i(\xi)}{\partial \xi^2} + \frac {1}{4}\,s_i\,
\left(\frac{\kappa}{l}\right)^2\,h_\mathrm{out}^{1/l}\, \xi^{1/l-2}\,
f_i(\xi) =0\, , 
\label{eq:sturm-liuv1}
\end{equation}
where $s_i$ or $k_i=s_i^2 \kappa^{2}
h_\mathrm{out}^{1/l}$ plays a role of
an eigenvalue, and the boundary condition $\partial f_i(\xi)/\partial \xi
=0$ at $\xi=\xi_\mathrm{out}$.  Changing to the free variable $x=\xi^{1/2l} =
(h/h_\mathrm{out})^{1/2l}$, we can write for a non-integer $l$
$$
f_i(x) = (k_i x)^l \, [\tilde A_i\, J_l(k_i\,x) + \tilde B_i\, J_{-l}(k_i\,x)]\, ,
$$
where $\tilde A_i$ and $\tilde B_i$ depend on the inner-boundary condition. According to the
Steklov theorem, the solution to (\ref{eq:problim_inh2}) can be obtained
in the form:
\begin{equation}
\tilde F(h,t) = \sum_{n=1}^{\infty} u_i(t) f_i(x)\, , 
\label{eq:steklov}
\end{equation}
where coefficients $u_i(t)$ are to be found from the inhomogeneous 
first-order differential equation over $t$. Let us substitute
(\ref{eq:steklov}) into (\ref{eq:sturm-liuv1}) and 
(\ref{eq:problim_inh2}), multiply by $f_n(x)$ and integrate over $x$
from 0 to $1$, using the orthogonality property of the eigenfunctions.
We arrive at 
\begin{equation}
\begin{split}
&\frac{{\partial u_n(t)}}{\partial t} + s_i\, u_n(t) = \Phi_n(t)\, , \\
&u_n\,|_{t=0}  = \phi_n \, ,
\end{split}
\label{eq:problim_inh3}
\end{equation}
with designations
\begin{equation}
\phi_n = \int\limits_0^1  \tilde F_0(x)\, f_n(x) \mathrm{d} x\, \Big/\,
\Vert f_n\Vert^2 ;~
  \Phi_n(t) = \int\limits_0^1  \Phi(h(x),t)\, f_n(x) \mathrm{d} x\,
\Big/ \,
\Vert f_n\Vert^2~ ,
\end{equation}
where $\Vert f_n\Vert^2$ is the norm of  eigenfunction $f_n$.
Solution to \eqref{eq:problim_inh3} is as follows
$$
u_i(t) = \phi_i \, e^{-s_i\,t} + 
\int\limits_0^t \Phi_i(\tau)\, e^{-s_i\, (t-\tau)} \, \mathrm{d} \tau\, .
$$
Let us write down the solution to (\ref{eq:problim_inh2}) for the case
$\dot M_\mathrm{out}=0$, that is, $\Phi(h,t)= 0$:
$$
\tilde F(x,t) = \sum_{i=1}^\infty \phi_i\, e^{-s_i\, t} \, f_i(x)\, .
$$
The last expression is equivalent to \eqref{eq:gen_sol_finite}.
In Sect.~\ref{sec:ev_fin_disk}, we find the coefficients $\phi_i=A_i$
and $B_i$ for the Dirac delta function as the initial condition and
express  $F(x,t)$ using the Green's function. In the analogy, 
the solution to the problem with $\Phi(h,t)\neq 0$ has the view 
\begin{equation}
\tilde F (x,t) = \int\limits_0^1  \tilde F_0(x_1)\, G(x,x_1,t)\,
\mathrm{d} x_1~ + \iint\limits_{0~0}^{t~~1}   \Phi(x_1,\tau) \,
G(x,x_1,t-\tau)\, \mathrm{d} x_1\, \mathrm{d} \tau\, ,
\label{eq:sol_nonzeroMdot}
\end{equation}
where $G(x,x_1,t)$ is given by (\ref{eq:green_zeroBi}) or
(\ref{eq:green_zeroAi}). 

Substituting (\ref{eq:sol_nonzeroMdot}) into (\ref{eq:substit}) and
using functions $\tilde F_0$ and $\Phi$ from (\ref{eq:problim_inh2}), we
finally arrive at (\ref{eq:sol_nonzeroMdot_full}). The coordinate
integral in its second term can be taken analytically; it involves the
Lommel functions and can be tabulated beforehand for the particular
values of $l$ and $k_i$ for the sake of the computational efficiency.
The second term of \eqref{eq:sol_nonzeroMdot_full} for $B_i=0$ can be
expressed as:
\begin{equation}
h_\mathrm{out}\, x^l  \sum_i J_l(k_i\,x) 
\int\limits_0^t \exp \left(-
\frac{t-\tau}{t_\mathrm{vis}}\, \frac{k_i^2}{2l}\right) \,
\left[ 4\, l\,L_1 \frac{\dot M_\mathrm{out}(\tau)}{t_\mathrm{vis}} - 
L_2\, \ddot M_\mathrm{out}(\tau) 
\right] \mathrm{d} \tau~,
\end{equation}
where
\begin{equation}
L_1 = \int\limits_0^1 x_1^{3\,l-1}\, J_l(k_i\,x_1) \, \mathrm{d} x_1 \, \Big /
J_l^2 (k_i);~
L_2 = \int\limits_0^1 x_1^{3\,l+1}\, J_l(k_i\,x_1) \, \mathrm{d} x_1 \, \Big /
J_l^2 (k_i)~.
\end{equation}


\end{document}